%% file: xs_Lya_Pkv7.tex
\documentclass[fleqn,usenatbib]{mnras}
\usepackage[T1]{fontenc}

\usepackage{graphicx}	
\usepackage{amsmath}	
\usepackage{amssymb}	
\usepackage{subfig}

\newcommand{\lya}{Ly-$\alpha$}
\newcommand{\lyb}{Ly-$\beta$}

\newcommand{\angstrom}{\textup{\AA}}
\newcommand{\Mpch}{h^{-1}\,\mathrm{Mpc}}
\newcommand{\skm}{\mathrm{km^{-1}\,s}}
\newcommand{\kms}{\mathrm{km\,s^{-1}}}

\newcommand{\be}{\begin{equation}}
\newcommand{\ee}{\end{equation}}

\usepackage{color}

\title[XQ-100: Lyman-alpha forest power spectrum]{The Lyman-alpha forest
  power spectrum from the XQ-100 Legacy Survey}

\author[V. Ir\v{s}i\v{c} et al.]{
Vid Ir\v{s}i\v{c}$^{1,2}$\thanks{E-mail: irsic@uw.edu (VI)},
Matteo Viel$^{3,4}$ \thanks{E-mail: viel@oats.inaf.it (MV)},
Trystyn A. M. Berg$^{5}$,
Valentina D'Odorico$^{3}$,
\newauthor
Martin G. Haehnelt$^{6}$,
Stefano Cristiani$^{3,4}$,
Guido Cupani$^{3}$,
Tae-Sun Kim$^{3}$,
\newauthor
Sebastian L{\'o}pez$^{7}$,
Sara Ellison$^{5}$,
George D. Becker$^{6,8}$,
Lise Christensen$^{9}$,
\newauthor
Kelly D. Denney$^{10}$,
G\'{a}bor Worseck$^{11}$ 
and James S. Bolton $^{12}$
\\
$^{1}$The Abdus Salam International Centre for Theoretical Physics,
Strada Costiera 11, I-34151 Trieste, Italy\\
$^{2}$University of Washington, Department of Astronomy, 3910 15th Ave
NE, WA 98195-1580 Seattle, USA\\
$^{3}$INAF - Osservatorio Astronomico di Trieste, Via G. B. Tiepolo
11, I-34143 Trieste, Italy\\
$^{4}$INFN - National Institute for Nuclear Physics, via Valerio 2,
I-34127 Trieste, Italy\\
$^{5}$Department of Physics and Astronomy, University of Victoria,
Victoria, BC V8P 1A1, Canada\\
$^{6}$ Institute of Astronomy and Kavli Institute of Cosmology, Madingley Road, Cambridge CB3 0HA, UK \\
$^{7}$Departamento de Astronom\'{i}a, Universidad de Chile, Casilla
36-D, Santiago, Chile\\
$^{8}$Space Telescope Science Institute, 3700 San Martin Drive,
Baltimore, MD 21218, USA\\
$^{9}$Dark Cosmology Centre, Niels Bohr Institute, University of
Copenhagen, Juliane Maries Vej 30, DK-2100 Copenhagen, Denmark\\
$^{10}$Department of Astronomy, The Ohio State University, 140 West
18th Avenue, Columbus, OH 43210, USA\\
$^{11}$Max-Planck-Institut f\"{u}r Astronomie, K\"{o}nigstuhl 17,
D-69117 Heidelberg, Germany \\
$^{12}$ School of Physics and Astronomy, University of Nottingham, University Park, Nottingham, NG7 2RD, UK\\
}

\date{Accepted XXX. Received YYY; in original form ZZZ}

\pubyear{2016}

\begin{document}
\label{firstpage}
\pagerange{\pageref{firstpage}--\pageref{lastpage}}
\maketitle

\begin{abstract}
We present the Lyman-$\alpha$ flux power spectrum measurements
of the XQ-100 
sample of quasar spectra obtained in the context of the European Southern
Observatory Large Programme "Quasars and their absorption lines: a
legacy survey of the high redshift universe with VLT/XSHOOTER". Using
$100$ quasar spectra with medium
resolution and signal-to-noise ratio we measure the power spectrum
over a range of redshifts $z = 3 - 4.2$ and over a range of
scales $k = 0.003 - 0.06\,\skm$. The
results agree well with the measurements of the
one-dimensional power spectrum found in the literature. The data
analysis used in this paper is based on the Fourier transform and has
been tested on synthetic data. Systematic and statistical uncertainties of our
measurements are estimated, with a 
total error (statistical and systematic) comparable to the one of the
BOSS data in the overlapping range of scales, and smaller by more than
$50\%$ for higher redshift bins ($z>3.6$) and small scales ($k > 0.01\,\skm$). 
The XQ-100 data set has the unique feature of having signal-to-noise ratios and resolution
intermediate between the two data sets that are typically used to perform cosmological
studies, i.e. BOSS and high-resolution spectra (e.g. UVES/VLT or HIRES).
More importantly, the measured flux power spectra span the high
redshift regime which is  usually more constraining for structure formation models.
\end{abstract}

\begin{keywords}
cosmology: observations -- (cosmology:) large-scale structure of the universe -- (galaxies:)
intergalactic medium -- methods: data analysis
\end{keywords}



\section{Introduction}
\label{sec:intro}

The absorption features blueward of the Lyman-$\alpha$ (\lya) emission
line in the spectra of high-redshift quasars (QSOs) are widely
used as biased tracers of the density fluctuations of a
photo-ionized warm intergalactic medium (IGM), and are collectively known as
the \lya\ forest (see \cite{meiksin09,mcquinn15} for recent reviews).

Although the first speculations and measurements were made almost $50$
years ago \citep{gunn65,lynds71}, the physical picture of the \lya\ forest was
established in the 1990s by a detailed comparison of analytic
calculations \citep{bi97,hui98,viel02} and numerical simulations
\citep{cen94,zhang95,miralda96,hernquist96,theuns98,theuns02}
with observed absorption spectra (e.g. \cite{kim04}).

In the last decade, a range of different statistics have been proposed \citep{schaye00,ricotti00,theuns00,theuns02,viel05,bolton08,lidz10,becker11,rudie12,garzilli12,bolton12,lee14,irsic13,boera14}, and
successfully used, that focused on specific aspects (e.g. targeting
cosmology, temperature of the IGM, etc.). However, the main quantity
of choice when comparing observations with the theoretical predictions
has become the one-dimensional flux power spectrum $P_F(k)$
\citep{croft99,croft02,kim04,viel04,mcdonald05,viel13wdm,palanque13}. This
is because the flux power spectrum is
tracing the actual fluctuations in the observed forest, making it easy to understand systematics and the
noise properties. The flux power spectrum also more cleanly
decouple the scales involved (e.g. fluctuations due to poor continuum
fitting are restricted to large scales).

Several measurements of the flux power spectrum have been performed in the last
two decades, ranging from measurements on a few ten high-resolution, high
signal-to-noise ratio QSO spectra
\citep{hires94,kim04,viel04,viel13wdm} to measurements on
many thousands of QSO spectra with poor resolution and signal-to-noise
\citep{sdss00,boss13,mcdonald05,palanque13}. Taken together, these
measurements cover over three orders of magnitude in scale  ($k =
0.001 - 0.1\,\skm$), however, they are either only centered on
large scales, or only on small scales, and no study  has
done a combined measurements of both.

In this paper we present a new set of measurements of the
one-dimensional $P_F(k)$ on an intermediate data-set: a hundred
QSO spectra with medium resolution ($\sim 10-20\,\kms$) and medium
signal-to-noise ratio ($S/N \sim 10-30$). The goal is to achieve
measurements of both large and small scales simultaneously and thus
provide a bridge between the traditionally used data-sets probing either large or
small scales.

The paper is structured as follows: in Sec.~\ref{sec:data_all} we
discuss the observational data used in our analysis, as well as the
synthetic data on which the data analysis procedure was tested. The
various steps of the data analysis are described in detail in
Sec.~\ref{sec:analysis}. The final results are presented in
Sec.~\ref{sec:results} and we conclude in Sec.~\ref{sec:conclusion}.

\section{Data \& Synthetic Data}
\label{sec:data_all}

\subsection{XQ-100 Sample}
\label{sec:data}

In this work we use 100 QSO spectra from the XQ-100 Legacy Survey
\citep{lopez16}, observed with the X-Shooter spectrograph on the
Very Large Telescope \citep{vernet11}. These 100 quasars span the
redshift range $3.51 < z < 4.55$.

We limit ourselves to spectra obtained from the UVB and VIS spectrograph arms (see
\cite{lopez16} for more details), since the
near-infrared spectral range gives us no information regarding the
\lya\ forest. For each QSO spectrum we merge the two spectral arms into one
spectrum by a simple method. We re-bin the spectra onto a fixed
wavelength grid with $\Delta \log_{10}{\lambda} = 3 \times 10^{-5}$
(with $\lambda$ in \AA),
which is the larger of the two bin sizes of the individual arms. In
the region where the arms overlap we perform weighted average of the
flux, continuum and resolution element. We have performed a test where
we treated each spectral arm as independent quasar observation and the
results showed that the effect of simple merging has negligible effect
on the flux power spectrum measurements, at least at the scales where
we are able to measure it.

Since the weighting is done using the optimal inverse variance
weights, any bad pixel that was determined to be so during pipeline
reduction analysis is thus down-weighted. However, the subsequent
merged spectra are also examined by eye if they make sense and don't
have any pixels that are obvious outliers. Using weighted merging of
the arms also ensures that the continuum transition from one arm to
the other is smooth. Whereas this introduces some false large scale
fluctuations in the continuum was not thoroughly explored, however any
such contributions would show up as excess of continuum power, which
we have investigated and verified it is very small (comparable to the
noise levels), see Sec.~\ref{sec:cont_errors}.

The resolution elements were taken to be constant per arm, with the
values of $20$ and $11\,\kms$ for UVB and VIS arms,
respectively.

The continuum used in our analysis is based on cubic spline fits and is described in more detail in
\cite{lopez16}.

After the spectral arms have been merged we perform additional cuts on
the data. First, we exclude pixels with negative or zero flux
errors as well as any bad pixels (with very negative flux of 
$f < -10^{-15}$, or as a flux over continuum level $f/C < -100$).

Second, we mask regions around Damped Lyman-$\alpha$ (DLA) systems
using the DLA sample provided by the survey team \citep{sanchez16}. We do not use
data within $1.5$ equivalent widths from the center of the DLA.

When measuring the flux power spectrum within the \lya\ forest we only
use the pixels within the $1045 - 1185\, \angstrom$ restframe wavelength
range of each QSO spectrum. This range is conservative in the sense that we
do not probe the absorption region close to the quasar \lya\ and
\lyb\ emission lines \citep{mcdonald05}. 

\subsection{Synthetic data}
\label{sec:mocks}

Our data analysis pipeline was tested with synthetic data that were generated
exclusively for this work. We want to generate a realistic flux field
with a QSO redshift distribution matching that of the observed data
sample. 

\begin{figure}
  \centering
  \includegraphics[width=1.0\linewidth]{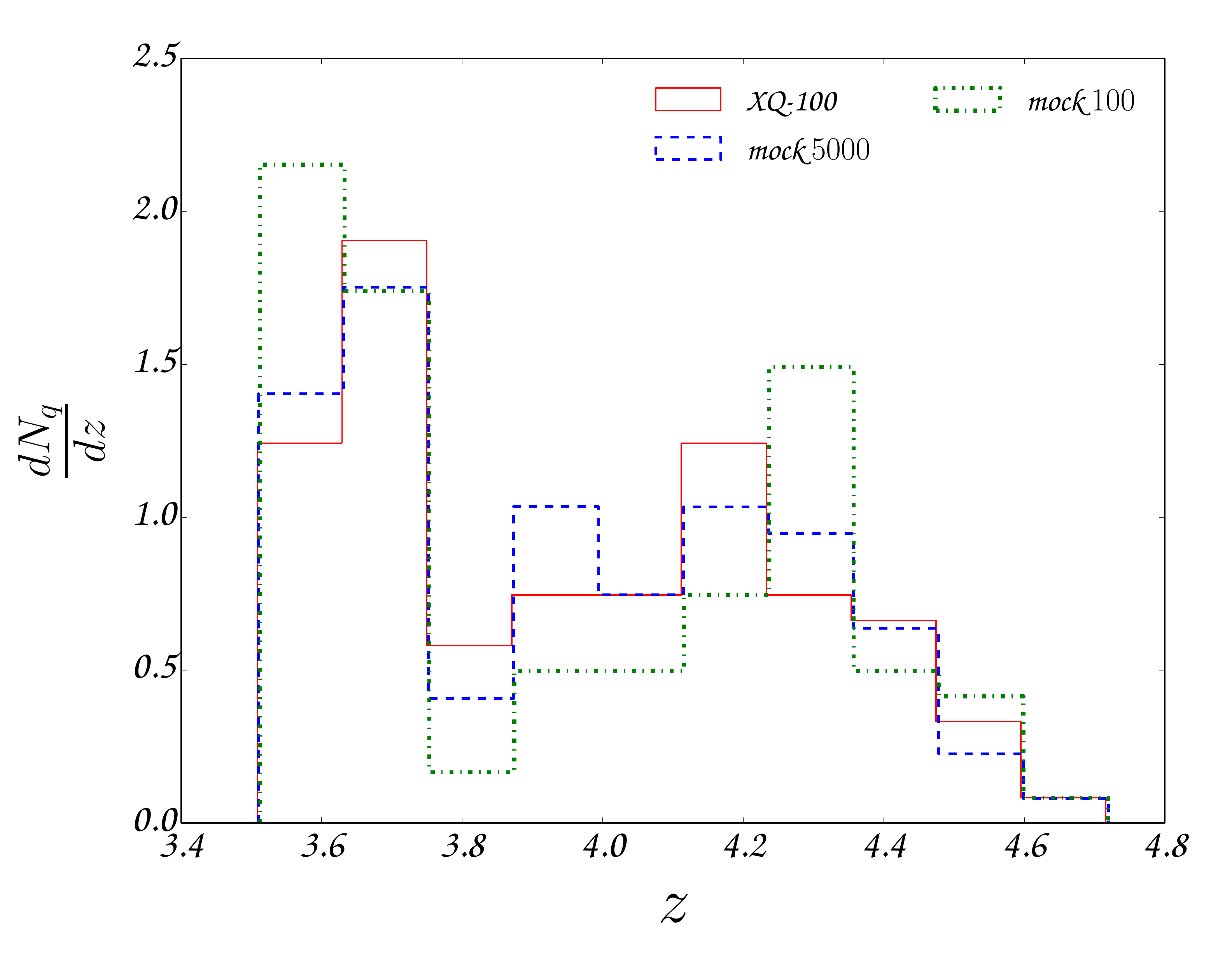}
  \caption{
    The QSO redshift distribution for XQ-100 data sample (red), for
    $N_Q=5000$ quasars of the synthetic data sample (blue), and for $N_Q=100$
    quasars of a synthetic data sub-sample (green).
  }
\label{fig:qso_zdist}
\end{figure}

First, we approximated the observed QSO redshift distribution by
binning the emission redshifts of the XQ-100 sample into $10$ redshift
bins, as shown in Fig.~\ref{fig:qso_zdist}. To generate synthetic
QSO sample we have drawn their redshifts from this
distribution. Figure~\ref{fig:qso_zdist} shows the distribution of
$5000$ and $100$ randomly drawn QSO redshifts from the distribution given by the data.

The various mock QSO catalogues used in this paper are presented in
Table~\ref{tb:mocks}.

\begin{table}
\centering
\begin{tabular}{|c|c|c|}
name & $N_Q$ & pixel size/resolution \\ \hline\hline
mock 5000 & 5000 & XQ-100 values \\ \hline
mock 100 & 100 & XQ-100 values \\ \hline
\end{tabular}
\caption{Different mock catalogues used in testing the data-analysis routine.}
\label{tb:mocks}
\end{table}

In the next step we want to produce flux spectra along the line of
sight of each QSO from the synthetic catalogues.
To this end we use a suite of high resolution hydro-dynamical
simulations of the intergalactic medium between redshifts 
$3 < z < 5$, with $2\times2048^3$ particles in a $40\,\Mpch$ box
size (PRACE: Sherwood simulations - \cite{bolton16}). The outputs were
produced with a redshift step of $\Delta z=0.1$ in a given redshift
range, in the form of an extracted optical depth along 
$5000$ randomly selected lines-of-sight.

For each  line-of-sight, and each redshift bin, the simulated optical
depth is given on a velocity grid ($\tau(v)$). 

First, we convert this
to a grid of wavelengths ($\lambda$), or equivalently \lya\ absorption
redshifts ($1+z = \lambda/\lambda_{\alpha}$), where $\lambda_\alpha$
stands for \lya\ line ($1215.67\;$\AA). The conversion is done so that
the mean absorption is assumed to happen at the
redshift bin of the simulation output ($z_s$)
\be
\lambda = \lambda_\alpha \left(1 + z_s\right) \sqrt{\frac{1 +
    \frac{v}{c}}{1 - \frac{v}{c}}},
\ee
where $v$ is the velocity coordinate along the line-of-sight within a
simulation box.
Since the length of the absorption spectrum along each  line-of-sight, at a given
redshift $z_s$ extends over the whole box size, and since the
cosmological simulations have periodic boundary conditions we make use
of that to extend the signal also to negative velocities by
periodically repeating the spectrum from a simulation box. Thus, for a redshift bin $z_s$ the
signal spans the redshift range of 
$z_s - \Delta z_s < z < z_s + \Delta z_s$, where $\Delta z_s$ is
simply the redshift length of the simulation box at a redshift
$z_s$. We choose to only repeat the periodic signal once, since in the
case of our simulations the redshift difference between each $z_s$ and
its neighbours is less than $2\Delta z_s$.

Second, we collect all the redshift outputs along each line-of-sight
into a single optical depth array. In principle the merging of the
simulation boxes at different redshifts can be
done using a variety of methods (e.g. weighted interpolation
between signals in neighbouring redshift bins). However we adopted
the simplest method and order them, one after the other, by increasing
simulation redshift, choosing simulation redshift bin with lower mean
redshift in the areas of overlap between two simulation outputs.

Such a construction allows us to have a line-of-sight extending over
many redshifts, and thus mimicking the observed spectrum. There are,
of course, a few shortcomings we would like to point out. 

Most
importantly, our basic ingredient is a spectrum extracted from a
numerical simulation with a given box size. Hence, we will only be
able to measure meaningful statistics on smaller scales. But we will
be able to do so for each redshift along a single line-of-sight.

Secondly, such a construction has rather discrete jumps in flux on the
border between regions from simulation outputs with different mean
redshift. The artifacts in a spectrum caused by such discrete jumps
can be avoided by using a more advanced technique of merging the
simulation outputs together along each line-of-sight, such as linear
(or higher order) interpolation. However, for our own tests on the
power spectrum, this did not play an important role, and thus we
settled for the simplest merging.

Thirdly, it is usually common to rescale the optical depth acquired
from simulations at a given redshift, so that the mean flux in that
redshift bin matches the observed one. Such re-scaling can
be viewed as a correction of the Ultra Violet background ionization rate from the
simulations to match the observed mean flux (due to degeneracy between
the two). The increase (or decrease) in the optical depth is usually
less than $20\%$. 

We performed a similar correction, but on the optical depth along the
entire constructed line-of-sight. The correction factor had a redshift
dependence, with redshift binning matching that of the simulation
output. The values were computed through iteration with the condition
that the mean flux computed along a specific line-of-sight matches one
from observations. For the purpose of testing the data analysis on
synthetic data it did not matter what exactly is the input observed
mean flux, as long as we recover it. We chose to use one given by
\cite{palanque13}. 

The last part in creating the synthetic data involved tailoring the
simulation output to a given survey specifications: QSO redshift
distribution, pixel-size, resolution and noise properties.

First, we assigned a
QSO emission redshift to each line-of-sight, thus specifying what
part of the redshift range falls in the \lya\ forest region for that
QSO spectrum. Quasars used in this procedure were determined by the
synthetic quasar catalogue. 

Each QSO spectrum was then rebinned with the same wavelength bin
size as in the XQ-100 observations 
($\Delta \log_{10}{\lambda} = 3 \times 10^{-5}$).

A convolution was performed on each spectrum with a Gaussian kernel
with resolution element of $33\;\kms$. Such a resolution
element is larger than the one in XQ-100 survey but for our purposes
of testing the data analysis procedure the exact number did not
matter.

In the end we also added noise to the spectrum. In principle adding
noise after the convolution with the resolution kernel only adds a
component that is flux-independent (e.g. read-out noise).
If the dominating contribution to the noise were flux dependent
(e.g.. Poisson noise) we could add it before convolving
with the resolution kernel. Both options were tested in the synthetic
data and subsequent data analysis, but for most of the tests presented
in the rest of the paper synthetic data has only flux-independent
noise component added.

\begin{figure}
  \centering
  \includegraphics[width=1.0\linewidth]{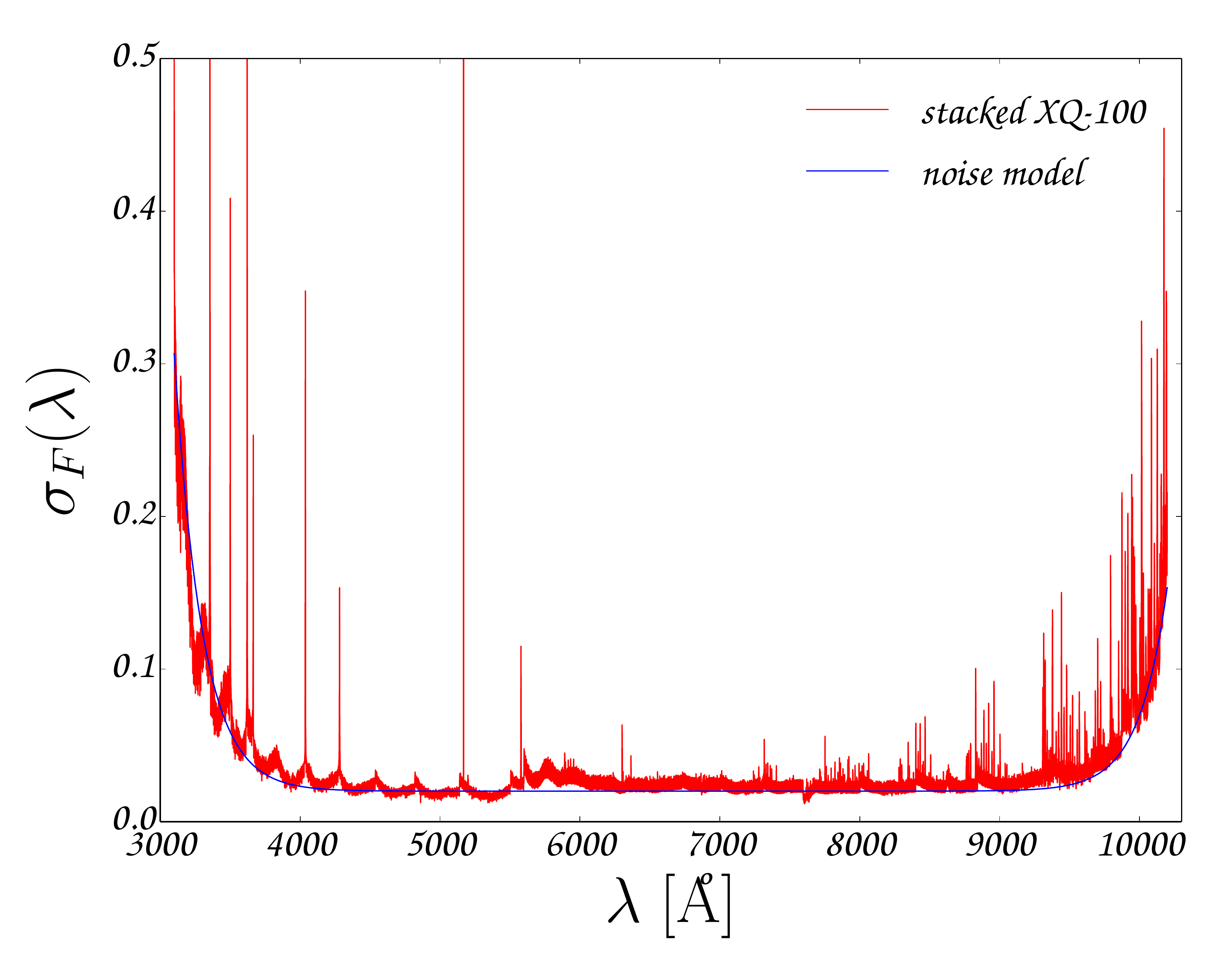}
  \caption{
    Averaged flux errors of the XQ-100 sample (in red) compared to the
    noise model used in constructing the synthetic spectra (in blue).
  }
\label{fig:noise_mocks}
\end{figure}

To make sure that our data-analysis routines correctly recognize and
subtract the noise we used a noise model that is comparable to 
averaging the flux errors of the actual XQ-100 data.
As shown in Fig.~\ref{fig:noise_mocks}, the noise has a slight
wavelength dependence towards the edges of the spectrum. Even though
the scatter is not negligible, it was not modeled in
the synthetic data. The function we used to describe the noise model
in the synthetic data is not a fit to the averaged flux errors. It is
just a simple closed form function that exhibits the same
large-scale wavelength-dependence behaviour. We found that for testing
purposes of this paper such a model was sufficient.

The very complicated flux error dependence comes from two instrumental
effects. First is that the flux error has a long wavelength mode
modulation, where it increases towards the edges of the observed
spectral range, which coincides with the edges of the CCD camera where
the pixel sensitivity is lower than in the middle of the CCD. This is
the effect we wanted to capture in the model of the flux errors since
a large mode fluctuations in real space of the flux errors might cause
sharp features in the Fourier space. We wanted to make sure we access
such a possibility on the mock data, and understand any potential
systematics it might cause. However, our error estimate did not show
any weird behaviour compared to having a constant value of flux error
with wavelength. Second effect on the observed flux error that causes
it to have a very complicated dependence was the small scale
modulation, which is caused by lower sensitivity at the overlapping higher Echelle
orders of the spectrograph. We did not model such a small scale
variation in our mock catalogues, since our error estimates on both
mock and real data would average over such small scales.

It should be noted that while the synthetic data in this paper were
designed for analysis of the flux power spectrum, they should be
applicable to other flux statistics as well.

\section{Data Analysis}
\label{sec:analysis}

In this section we describe the steps taken in the data analysis
procedure. The same strategy was adopted for both real and synthetic
data in order to check for any systematic effect arising due to the
analysis itself.

The bulk of the analysis consists of the Fourier transforms of the
input spectra, which is a method that has been used extensively
before, on similar data sets
\citep{croft99,croft02,kim04,viel04,viel13wdm}. 
This method is used to measure the flux
power spectrum of the \lya\ forest. The measurements, of both real and synthetic
data, are in $7$ $z-$bins ($z = 3.0 - 4.2$ with step 
$\Delta z = 0.2$) and $19$ $k-$bins ($k = 0.003 - 0.06\;\skm$, linearly
binned with step $\Delta k = 3 \times 10^{-3} \;\skm$).

\subsection{Continuum}
\label{sec:continuum}
Using the provided continuum fits for each QSO spectrum ($C$), we
  first divided the continuum of the XQ-100 spectrum measurement ($f$). While we tested the
  robustness of the results by using different continuum models, we
  opted in the end for the official XQ-100 continuum fits described in
  \cite{lopez16}. We did
  not fit the continuum at the same time as the mean flux or the power
  spectrum. In the synthetic data, the continuum was modeled as a
  constant equal to unity.
\subsection{Redshift sub-samples}
\label{sec:chunks}
For each line-of-sight we split the data into separate
  sub-samples ($z$-bins) by measured redshift. Each pixel is assigned an absorption redshift
  which determines the redshift of the sub-sample it falls into. We
  perform this step, so that the Fourier Transform used for the power
  spectrum analysis is performed on the level of $z$-bins and not
  on the whole line-of-sight. This is foremost much easier to handle,
  since the scales of different mean redshifts are not mixed
  together. It is also convenient to measure the power within a
  redshift bin where the variation in wavelength is described by a
  velocity coordinate only. This is an approximation, since measuring
  flux along a photons' path gives a relation between redshift and
  proper coordinate (or equivalently velocity coordinate). However the
  corrections are very small when measuring \lya\ power spectrum
  \citep{mcdonald06,irsic15}.
\subsection{Mean Flux}
\label{sec:meanF}
We perform an un-weighted average of the flux to
  obtain an estimate of the mean flux (${\bar F} = \langle F\rangle =
  \langle f/C\rangle$). A sample average
  gives us an unbiased estimator of the true value, but underestimates
  the error on the average. To perform the unbiased weighted average
  the full variance would have to be known (which is the sum of the
  error flux variance and variance due to cosmic
  fluctuations). However, the cosmic variance is not known at this
  stage in the data analysis. One option would be to measure the mean flux
  and its variance together through a likelihood based iteration
  scheme, or compute the variance from the measured power spectrum. We
  opted for the latter, and simpler method. 
\subsection{Flux Power Spectrum}
\label{sec:Pk}
For each line-of-sight, and each $z$-bin we
  perform Fourier Transform on a flux fluctuation field ($\delta_F
  = F/{\bar F} - 1$). The flux power estimator is then given as a sum of the
  squared Fourier coefficients over all the pixels in all the $z$-bins
  along all the lines-of-sight that contribute to the measured
  $(k,z)$ bin:
\be
{\hat P}_{\text{tot}}(k_i,z_j) = \frac{1}{N_{ij}} \sum_{n,m} |\delta_F(k_n,z_m)|^2 \delta_D(k_i
- k_n) \delta_D(z_j - z_m),
\ee
where $N_{ij}$ represents the number of pixels contributing to the bin
$(k_i,z_j)$. The sum goes over all the pixel pair configurations with
a wave number $k_n$ and redshift $z_m$. We have denoted the Dirac delta
function as $\delta_D$.

At this point we also correct the result for the effects of finite
pixel width and resolution element. Deconvolution of the flux
fluctuation field translates into simple division in the Fourier
space, thus
\be
\delta_F(k_n,z_m) =
\frac{\delta_F^{(measured)}(k_n,z_m)}{W^2(k_n;p_{n,m},R_{n,m})},
\ee
where $p_{n,m}$ is pixel width of pixel corresponding to bin ($k_n,z_m$) and $R_{n,m}$ is resolution
element of the same pixel. Both $p$ and $R$ are in velocity
units. The pixel width $p$ is constant in both our data sets ($p = c
\Delta \log_{10}{\lambda}$, with $\lambda$ in \AA), whereas the resolution element can vary and
is given for each pixel. In the synthetic data set, $R$ is constant and
equal to $33\;\kms$ but in the real data set it varies
between $11$ and $20\;\kms$ due to different resolutions in
different spectral arms.

The de-convolution kernel in Fourier space, $W(k;p,R)$,  is a product
of a Gaussian (Gaussian smoothing of the resolution element) and a
Fourier transform of a square function (pixel width):
\be
W(k;p,R) = e^{-\frac{1}{2}k^2 R^2} \frac{\sin^2(\frac{k p}{2})}{(\frac{k
  p}{2})^2}.
\ee

\subsection{Noise power}
\label{sec:noise}
In the subsection above we have explained how the total flux power
spectrum is evaluated. However, this power describes both fluctuations
due to noise and the cosmological signal we are interested in. It is
fair to assume that noise is uncorrelated with the cosmological
signal, and thus it can be removed at the power spectrum level:
\be
P_F(k,z) = P_{\text{tot}}(k,z) - P_N(k,z).
\ee

We estimate the noise power by assuming that the $P_N(k,z)$ can be
treated as constant in $k$, and its normalization for each redshift
can be obtained through the variance of the flux errors as a function
of redshift. To that end we compute the estimate of the flux error
variance, at the step when we compute the mean flux
\be
\sigma_N^2(z_j) = \sum_i \frac{\sigma_F^2(\lambda(z_i))}{M_j},
\ee
where $M_j$ is the number of pixels that correspond to a redshift bin
$z_j$. The noise power is then given by
\be
{\bar F}^2(z) \sigma_N^2(z) = \frac{1}{\pi} \int_0^\infty P_N(k,z) dk \approx
\frac{1}{\pi} P_N(z) \left(k_{\text{max}} - k_{\text{min}} \right),
\label{eq:noise_power}
\ee
where $k_{\text{min}} = 0$ for our choice of binning and
$k_{\text{max}}$ is equal to Nyquist scale, which is the largest
independent scale we measure through our Fourier Transform
analysis. 

The estimate obtained through the above relation is used in our data
analysis as the noise power. This method has been tested on synthetic
data (see next Section) and provides satisfactory results. 

\section{Results}
\label{sec:results}

\begin{figure}
  \centering
  \includegraphics[width=1.0\linewidth]{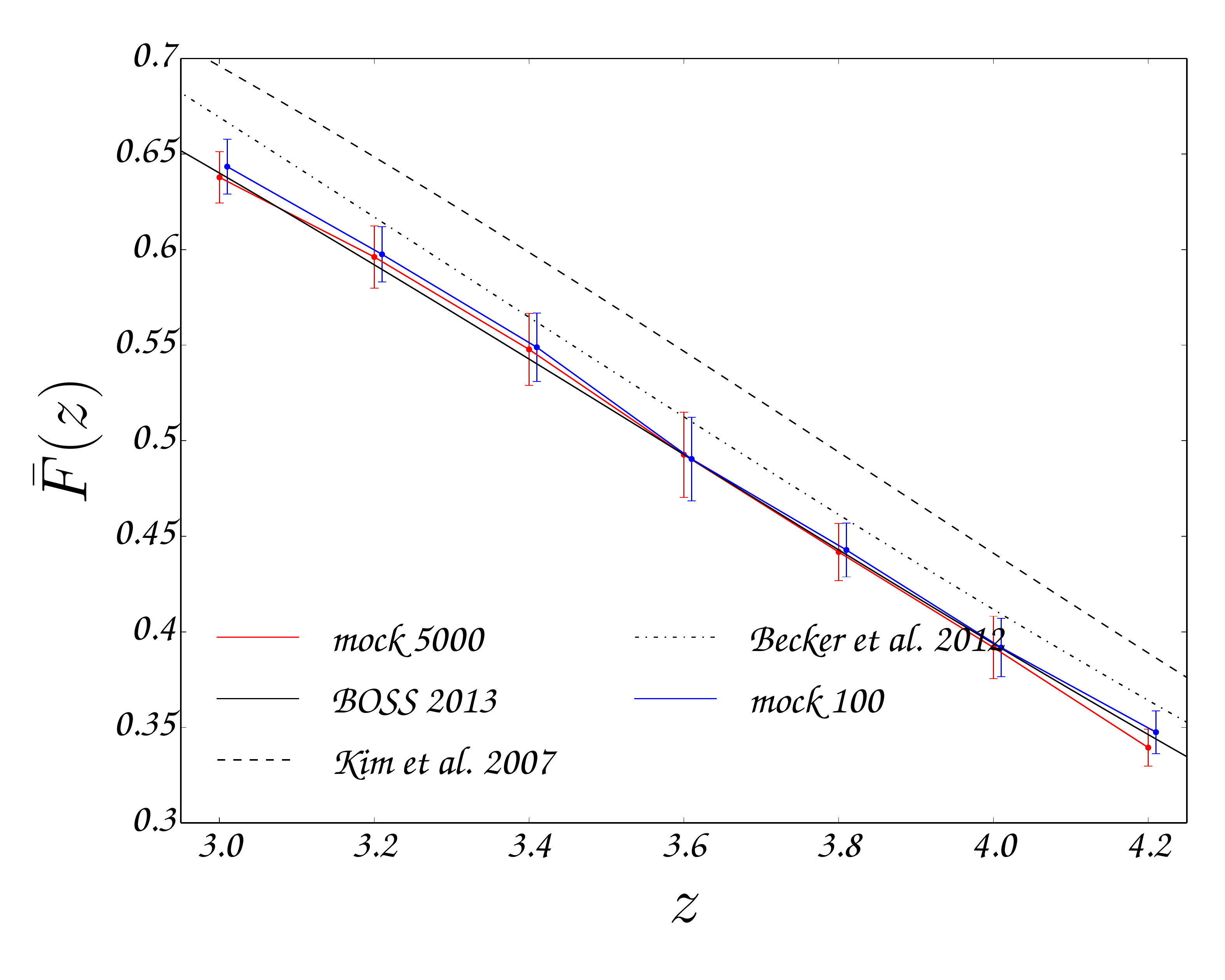}
  \caption{
    The mean transmitted flux as obtained from the synthetic data. In
    red we show results using $5000$ mock QSO spectra and using $100$
    mock QSO spectra (in blue). In black we plot the standard
    observational results by \protect\cite{kim07} (dashed), \protect\cite{becker12}
    (dot-dashed) and \protect\cite{palanque13} (full line). The input to the mocks was the
    BOSS mean flux. The data points are shifted in redshift (by
    $0.01$) to be readily distinguishable. 
  }
\label{fig:meanF_mocks}
\end{figure}

In this section we present the results of the data analysis procedure
presented in this paper. First we show the results and tests of
various methods and approximations used in the analysis of the
synthetic data. We then show the main results of this paper, performed
on the XQ-100 sample of QSO spectra. In the last subsection we
discuss the way to obtain the estimate of the errors on the flux power
spectrum bins.

\subsection{Power spectrum results on synthetic data}
\label{sec:results_mocks}

First we apply the data analysis procedure to the synthetic catalogue
$5000$ QSO spectra in order to test for possible
systematic effects in our analysis. By using a larger number of QSO spectra we hope to
beat down the statistical fluctuations and proclaim the deviations
that remain as systematic errors. 

The measurements of the mean flux
on synthetic data are presented in Fig.~\ref{fig:meanF_mocks}. The
input mean flux with which we have calibrated the simulation outputs
is plotted in full black line (BOSS 2013 - \cite{palanque13}). Red points with
error-bars are measurement from the data analysis procedure presented
in this paper. The results agree well with the input
version and suggests no important systematic effects are present in
this measurement. The analysis was also repeated on a synthetic
catalogue with only $100$ QSOs. The results are plotted in green in
Fig.~\ref{fig:meanF_mocks}, and agree well with the $5000$ QSO spectra
sample. Note however that the error-bars are very similar, and that is
because they are dominated by the variance of flux fluctuations.
As a comparison, Fig.~\ref{fig:meanF_mocks} also
shows observed flux from two other surveys (\cite{kim07} and \cite{becker12,viel13wdm})
on a different sample of measured real data spectra. 

\begin{figure}
  \centering
  \includegraphics[width=1.0\linewidth]{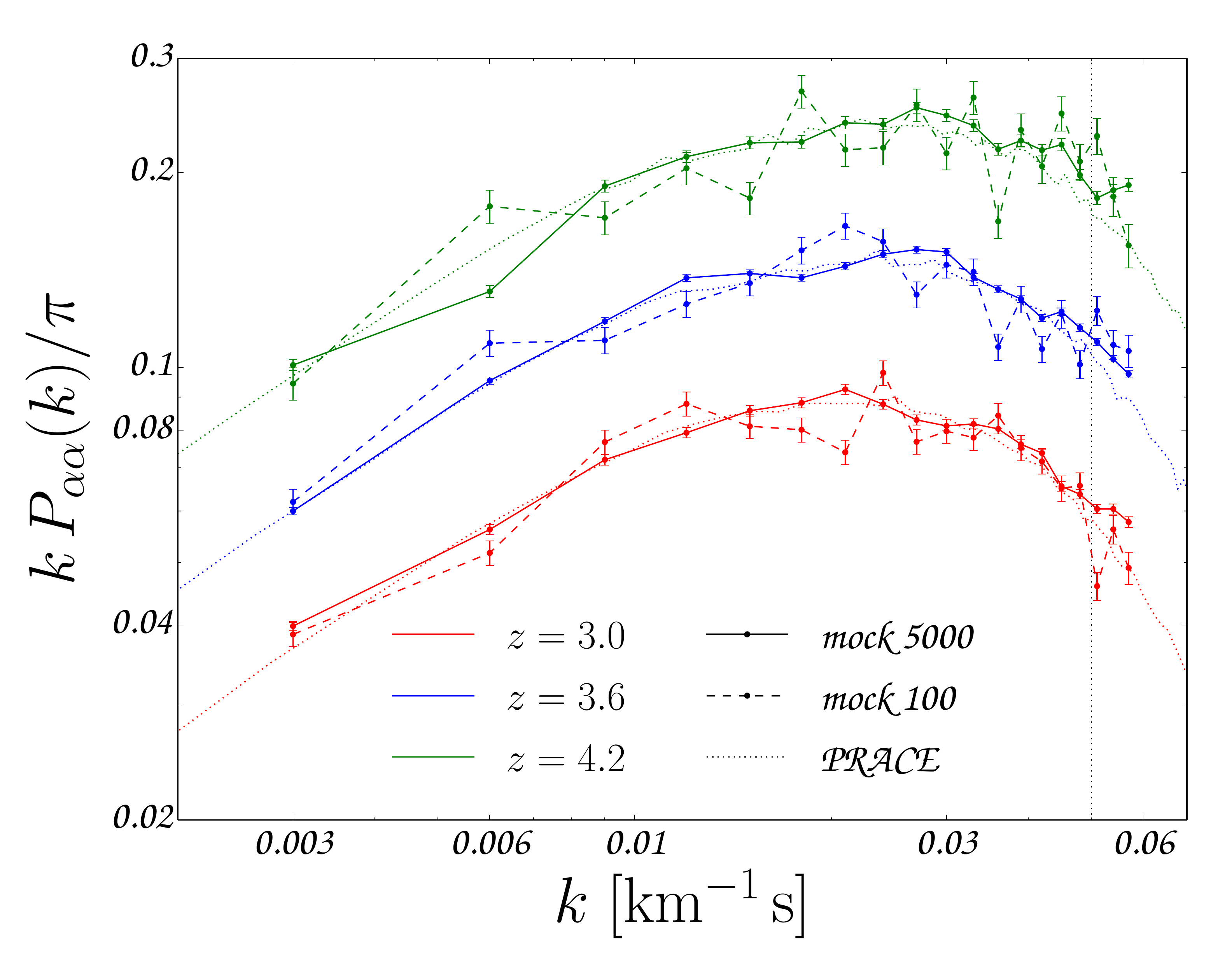}
  \caption{
    The flux power spectrum measured on the synthetic data for
    $5000$ (full lines) and $100$ (dashed lines) QSO spectra is shown. The three
    colours correspond to three (out of $7$ measured) redshift bins: 
    $z=3.0$ (red), $z=3.6$ (blue) and $z=4.2$ (green). The dotted
    lines correspond to the power spectrum extracted from a simulation
    at that redshift. The error-bars
    are evaluated using bootstrap method (see Sec.~\ref{sec:covmat} for details).
  }
\label{fig:Pk_mocks}
\end{figure}

Next, the data analysis was tested on the measurements of the flux power
spectrum. Fig.~\ref{fig:Pk_mocks} shows the results as a function of
scale ($k$) for three redshift bins ($z=3.0$ - red, $z=3.6$ - blue and
$z=4.2$ - green). The full lines represent the measurements performed
on the synthetically generated spectra as described in
Sec.~\ref{sec:mocks}. For comparison we show the flux power spectrum
obtained by measuring it directly on the simulation output at the
specified redshifts (using 5000 lines of sight), without going through the construction procedure
of the synthetic data (dotted lines). The departures from the input power spectrum
at large scales are due to insufficient number of lines-of-sight
probing those scales. This is apparent from looking at the dashed-line
in Fig.~\ref{fig:Pk_mocks} where the same analysis is performed on
only $100$ QSO spectra.

\begin{figure}
  \centering
  \includegraphics[width=1.0\linewidth]{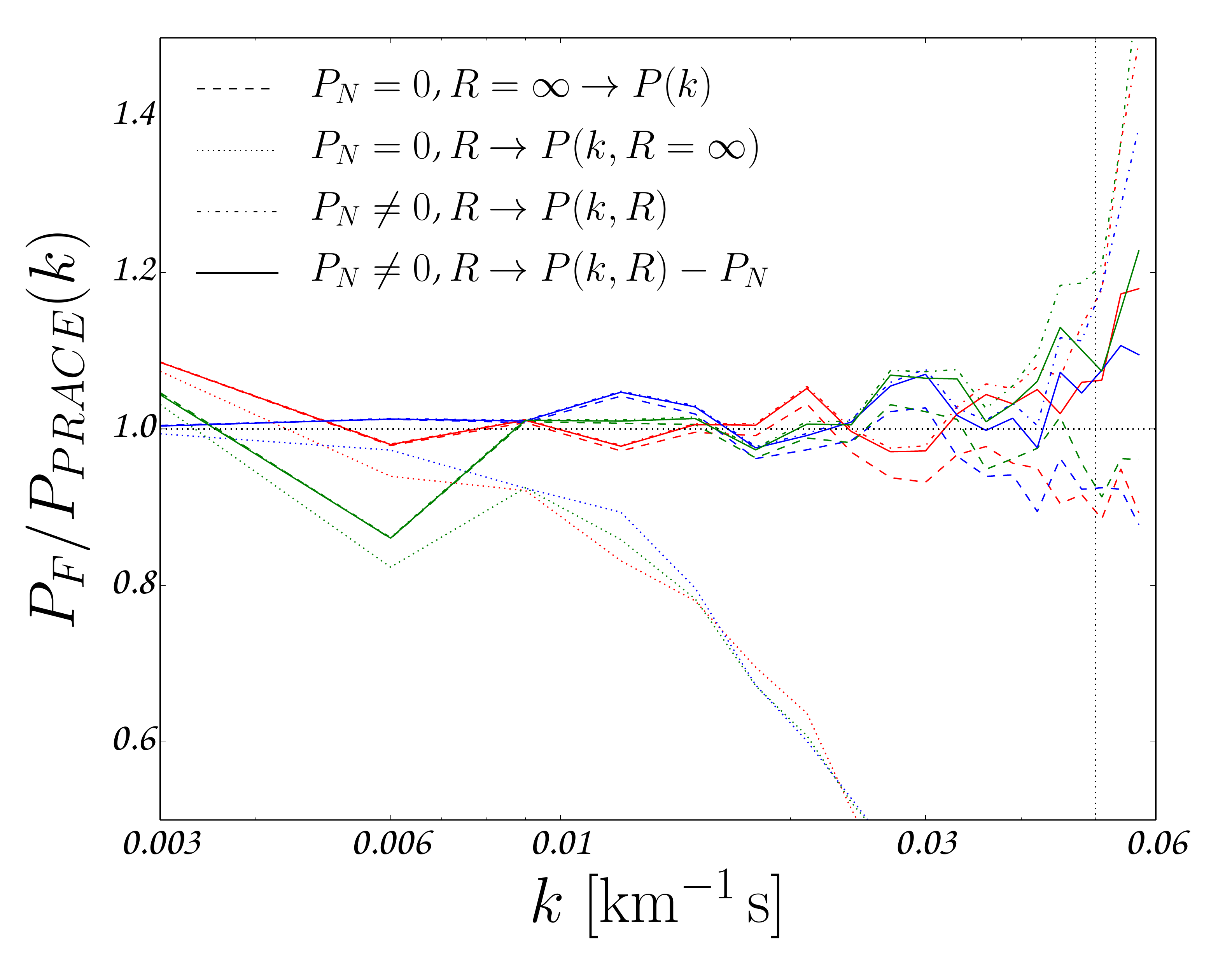}
  \caption{
    The ratio between flux power spectrum measured from the
    synthetic data (mock 5000) and the input simulation power
    spectrum is reported. The colours again correspond to three redshift bins (red
    - $z=3.0$, blue - $z=3.6$, green - $z=4.2$). Different line styles
    correspond to different assumptions when generating synthetic data
    as well as different data analysis steps taken: $P_N$ - whether
    noise is added to the synthetic data, $R$ - whether
    resolution/pixel width were added; $P(k,R)$ - whether in the data
    analysis resolution was corrected, and $-P_N$ whether noise was
    subtracted (see text for details).
  }
\label{fig:Pk_mocks_tests}
\end{figure}

However, there are still some fluctuations present at smaller scales
that persist even when increasing the number of QSO spectra in our
analysis of the synthetic data. Fig.~\ref{fig:Pk_mocks_tests} shows in
greater detail the ratio between the recovered flux power from the
synthetic data and simulation power spectrum at those redshifts. The
three colours still represent three redshift bins, but different line
style show different tests done in either the construction of the mock
data or the data analysis procedure. 

The
dashed coloured lines (Fig.~\ref{fig:Pk_mocks_tests}) show the effect of not correcting for the pixel
width. The lines show the recovered power spectrum from the
mocks, where no noise has been added ($P_N = 0$) and no convolution
with the resolution element has been performed performed ($R = \infty$). No corrections to noise,
resolution or pixel width were added when extracting the flux power
from the mocks. The ratio is different from
unity because in the synthetic data the spectra were rebinned using
XQ-100 wavelength bin size, while the flux power spectrum from
simulations was computed using much finer binning.

The dotted lines (Fig.~\ref{fig:Pk_mocks_tests}) shows the effect of not correcting for the resolution
element. The lines show the recovered power spectrum from the mocks
with no noise ($P_N = 0$), but spectra were convolved with a
Gaussian kernel with a resolution element $R$ (see
Sec.~\ref{sec:mocks}). However, no correction to the resolution was
made in the data analysis. Comparing with dashed lines, properly
correcting for the resolution has much bigger impact on the recovered
flux power than correcting for the pixel width.

Additional tests were performed, where both noise ($P_N \neq 0$) and
resolution ($R$) were added to the synthetic data, and while the data
analysis corrected for the resolution element, no correction to the
noise was added (dot-dashed coloured lines in
Fig.~\ref{fig:Pk_mocks_tests}). Not correcting for the noise clearly
introduces spurious power on small scales which increases rapidly,
while large scales remains unaffected.

\begin{figure}
  \centering
  \includegraphics[width=1.0\linewidth]{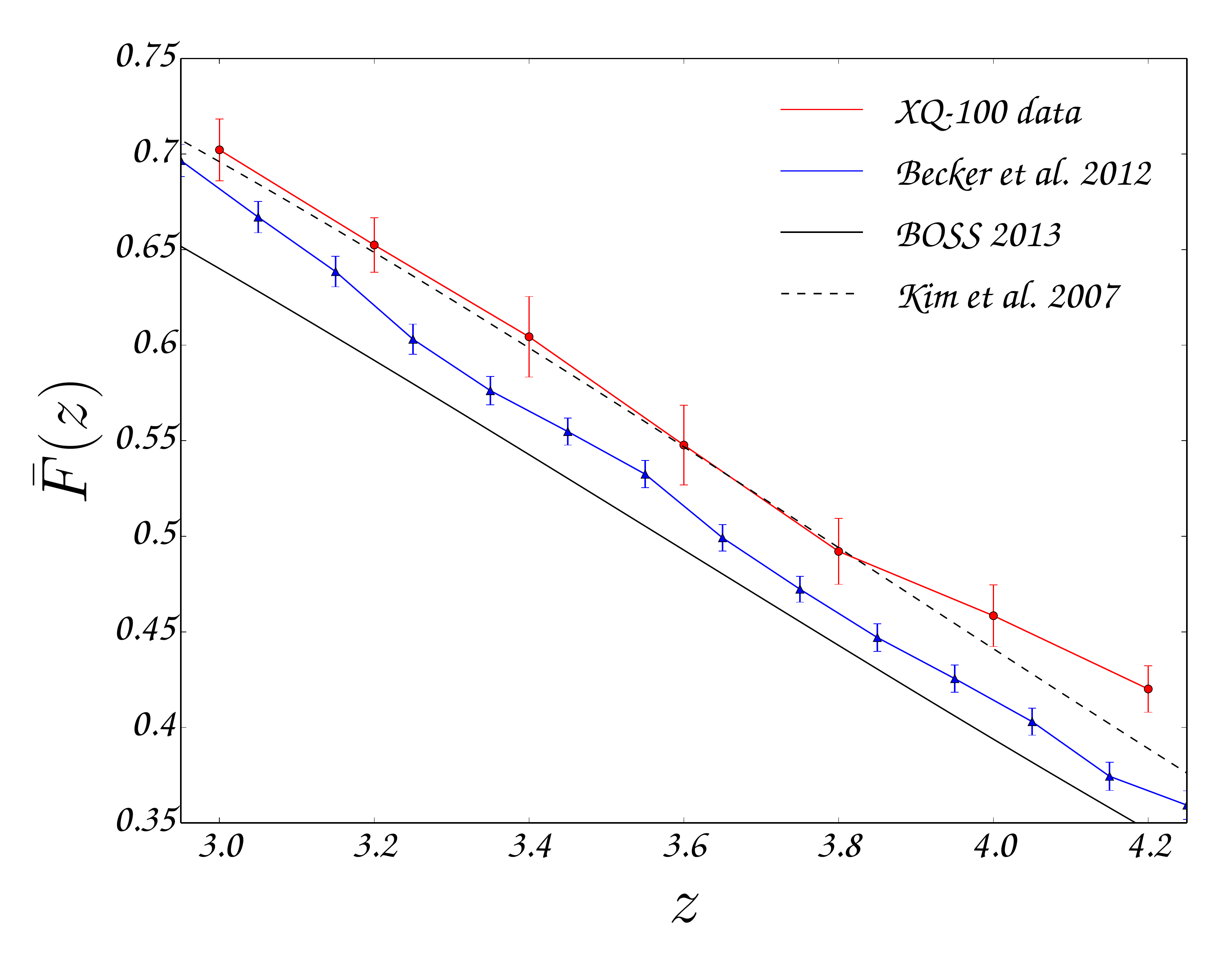}
  \caption{
    The mean transmitted flux measured on the XQ-100 data sample (red points) using
    the data analysis and cuts as described in Sec.~\ref{sec:analysis}
    and ~\ref{sec:data}. As a comparison we also plot results for mean
    flux from \protect\cite{palanque13} (full black line) and extrapolated
    values from \protect\cite{kim07} (dashed black line). The error bars on the mean flux
    were taken to be from the bootstrap covariance matrix. We also
    compare our results to the mean flux measurements by
    \protect\cite{becker12} (blue points). The difference comes from different
    continua estimation (see text for details).
  }
\label{fig:meanF_data}
\end{figure}

\begin{figure*}
  \centering
  \includegraphics[width=1.0\linewidth]{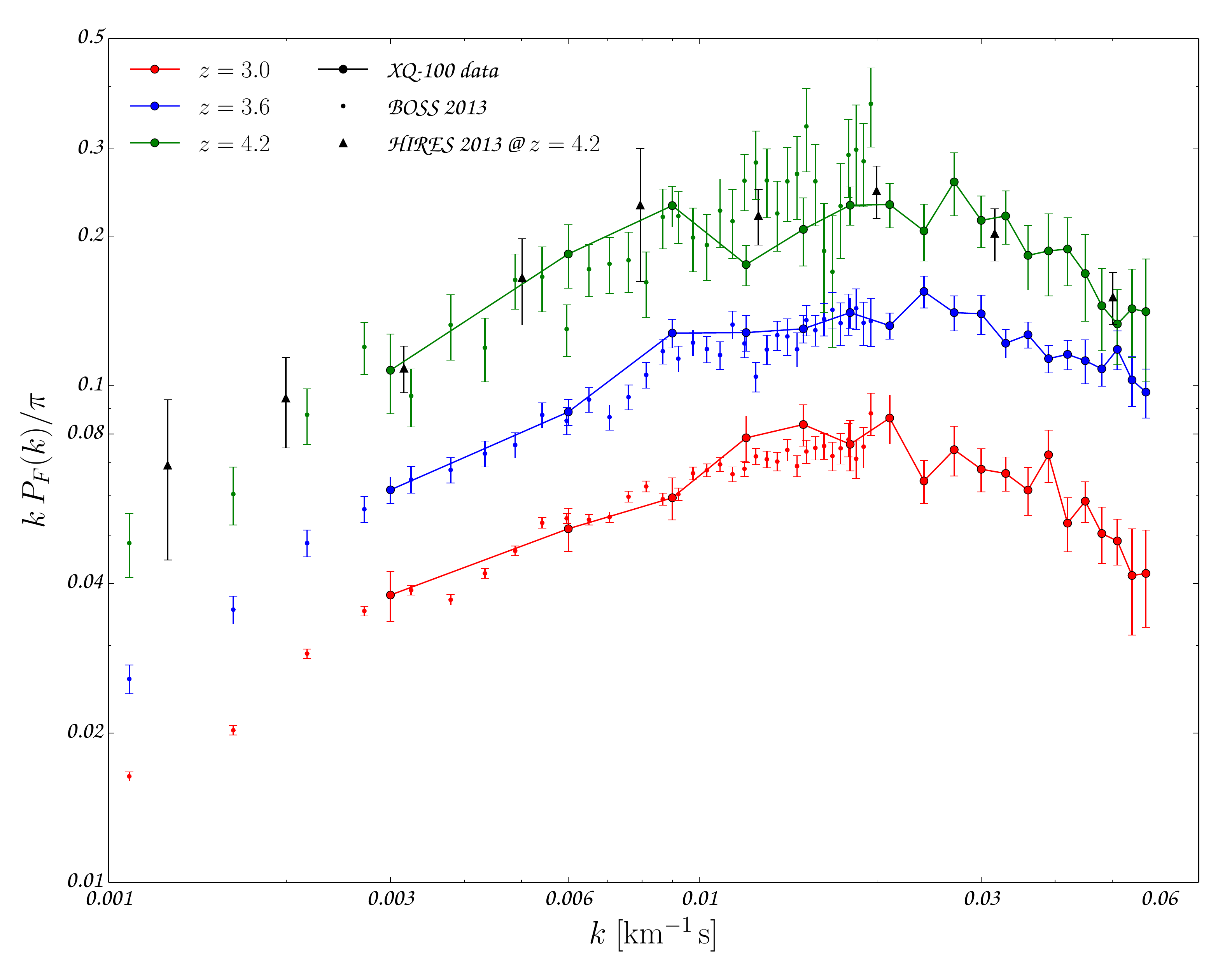}
  \caption{
    The flux power spectrum measurements of the XQ-100 data
    sample (circles). Full data analysis procedure described in
    Sec.~\ref{sec:analysis} was applied, as well as all the cuts to
    the data presented in Sec.~\ref{sec:data}. We have also subtracted
    the metal power spectrum (see Sec.~\ref{sec:metal_power}). The
    error-bars used in this plot are a squared sum of both statistical
    errors (from bootstrap matrix estimation) and systematic errors
    (see Sec.~\ref{sec:results_mocks}). As a comparison measurements
    from \protect\cite{palanque13} (dots) and \protect\cite{viel13wdm}
    (black triangles) are also plotted.
  }
\label{fig:pk_data}
\end{figure*}

The last test (full lines in Fig.~\ref{fig:Pk_mocks_tests})
shows the effect of correcting the resolution element with slightly
wrong value. We assume that our knowledge of the (synthetic) data
resolution element is of the order of few $\kms$ (or roughly
$10\%$). Both noise and XQ-100 pixel width were used in the
construction of the mock sample, and both were as well corrected for
in the power spectrum estimation. However the effect of misestimating
the resolution element translates into wrong power spectrum recovery
on small scales. Deviations of the flux power spectrum on small
scales ($k \sim 0.05\,\skm$) are of order of $5-10\%$. 

On large scales tests agree nearly perfectly with each
other, which indicates that the fluctuations there are
specific to the data-set not the data-analysis routine, and thus of
statistical nature. However on smaller scales the difference to the
simulation power are interpreted as correcting for slightly wrong
values of resolution element or pixel width (where resolution carries
more weight). The differences are again of the
same order of magnitude ($5-10\%$) at the small scale end of our
measurements.
Additional cause of these differences might be that no correction has been
made in the analysis regarding the aliasing of small scales
approaching Nyquist scale.
To account for these systematic effects in our data analysis we use
the results shown in full lines in Fig.~\ref{fig:Pk_mocks_tests} to
determine the systematic errors. The absolute difference between the
models shown in Fig.~\ref{fig:Pk_mocks_tests} (full lines), and the
reference line of unity was used as a systematic error standard
deviation.

\subsection{Lyman alpha flux power spectrum from XQ-100 sample}

This section contains the main results of the data analysis of the
XQ-100 data sample. First we present the measurements of the mean
transmitted flux as a function of redshift, in
Fig.~\ref{fig:meanF_data}. The error bars were obtained using the
method described in Sec.~\ref{sec:meanF}. As a comparison we also plot
mean flux fitting formulas from \cite{palanque13} and \cite{kim07}. The mean
flux measurements of XQ-100 data agree well with \cite{kim07} up to
redshift around $3.8$. 
The line for \cite{kim07} plotted in
this paper is in fact an extrapolation of the fitting formula
performed on lower redshift QSO spectra ($z < 3$). However comparing it to
our results, it seems to be valid even at higher redshifts. The
difference in mean flux normalization between our results and
\cite{palanque13} is probably due to different continuum fitting
procedures. 
In Fig.~\ref{fig:meanF_data} we also compare to the results of the
mean transmitted flux of \cite{becker12}. The difference is mainly due
to different continuum fitting. Moreover the results by
\cite{becker12} presented in this paper were rescaled to match
lower redshift measurements by \cite{faucher08}. Our data lacks the
sufficiently low redshifts ($z = 2 - 2.5$) to be used as rescaling of
the results by \cite{becker12}.

\begin{figure*}
  \centering
  \begin{tabular}{cc}
    \subfloat[$k$-dependence]{
      \includegraphics[width=0.5\linewidth]{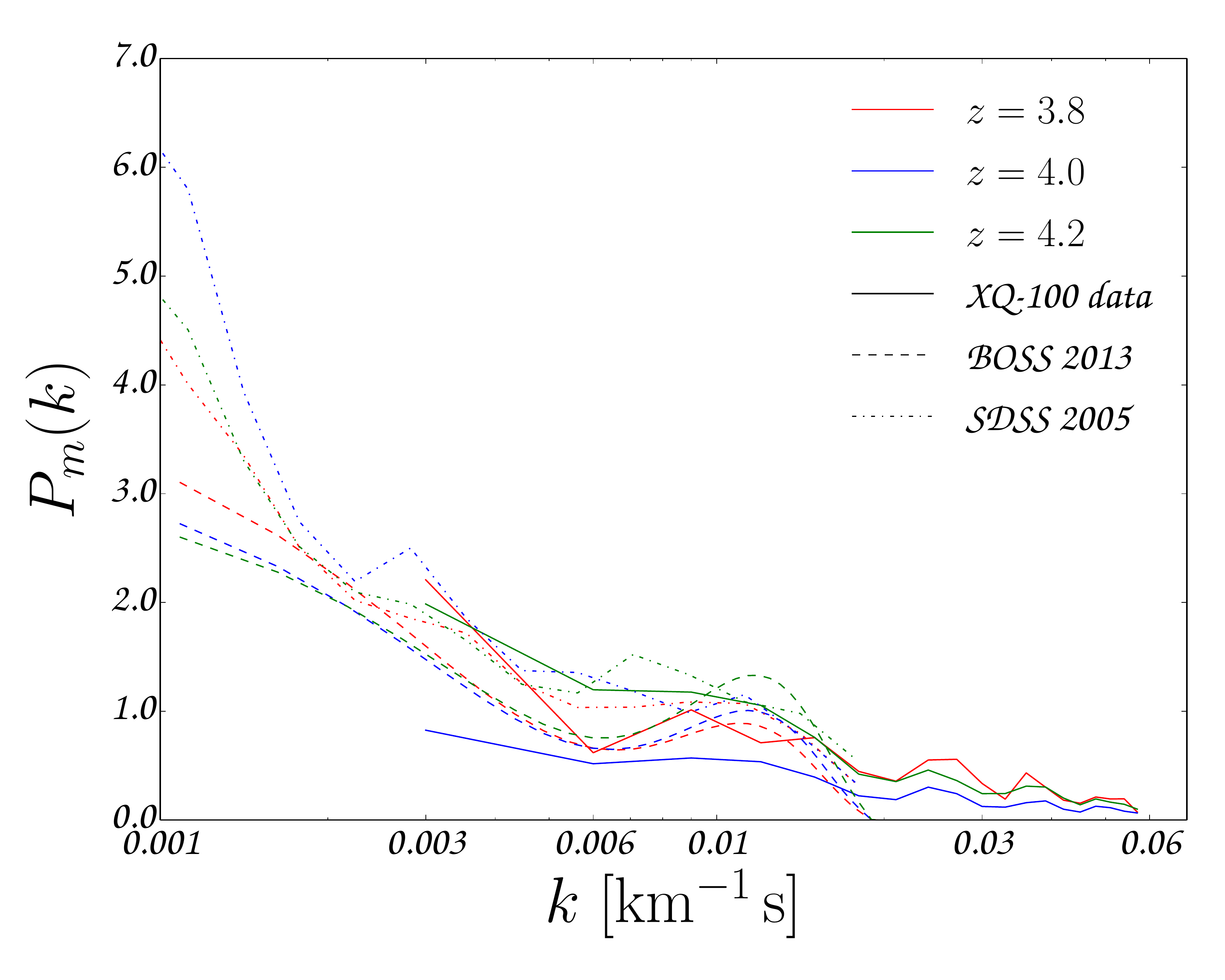}
      \label{fig:Pk_metal}
    } &

    \subfloat[$z$-dependence]{
      \includegraphics[width=0.5\linewidth]{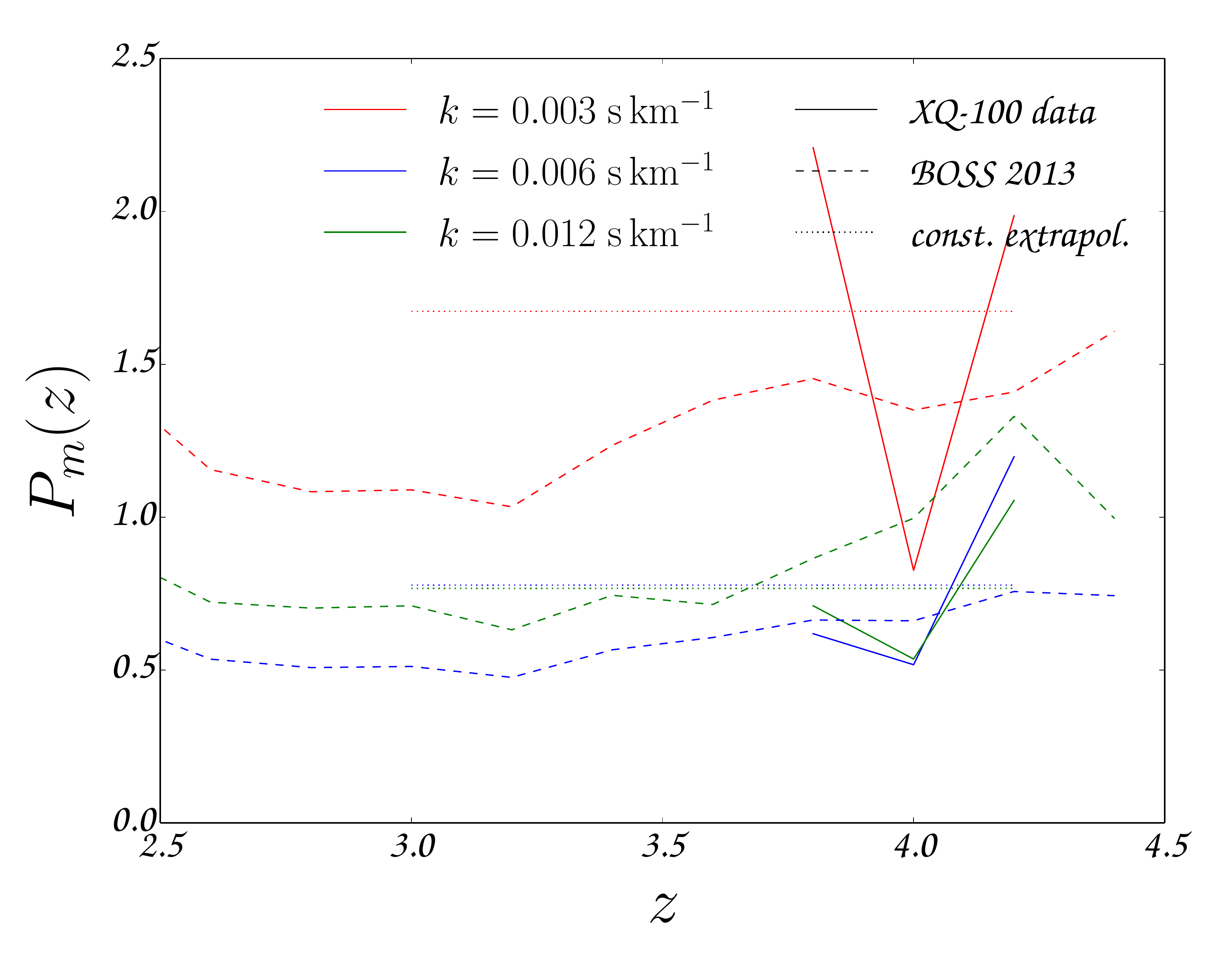}
      \label{fig:Pk_metal_zdep}
    } \\
  \end{tabular}
  \caption{
    {\it Left}: The metal power spectrum measured in the restframe redshift range
    of $1268 - 1380 \,\angstrom$ for three redshift bins:
    $z=3.8$ (red), $z=4.0$ (blue) and $z=4.2$ (green). Measurements of
    metal power spectrum by \protect\cite{mcdonald05} (dot-dashed lines) and
    \protect\cite{palanque13} (dashed lines) are also plotted. {\it Right}:
    The measurements of the metal power spectrum as a function of redshift, for three
    different $k$-modes: $k=0.003\,\skm$ (red), $k=0.006\,\skm$ (blue)
    and $k=0.012\,\skm$ (green). The dashed lines show the result by
    \protect\cite{palanque13}. Previous measurements of metal power
    (dashed lines) indicate that the
    redshift dependence can be approximated as roughly constant for
    each $k$-mode. Dotted lines show our result of such an
    approximation, which is also used to extrapolate $P_m(k,z)$ to
    lower redshift bins.
  }

\label{fig:Pk_metal_ab}
\end{figure*}

The most important result of our paper is present in
Fig.~\ref{fig:pk_data}. The figure shows the flux power spectrum,
measured on the XQ-100 sample of QSO spectra, as a function of
scale for three redshift bins from our analysis. All the steps from
the data analysis procedure were performed in order to obtain the flux power
values presented in this plot (see Table~\ref{tb:pk_data} for full
sample of measurements). We have also subtracted the metal power
spectrum, measured within the same data sample and extrapolated to
lower redshifts (see Sec.~\ref{sec:metal_power}).
As a comparison the
measurements of the BOSS 2013 analysis are also plotted
\citep{palanque13} as well as overlapping redshift from high-redshift
measurements \citep{becker12}.
Since XQ-100 data sample only has $100$ QSO spectra, the flux
power cannot be measured at scales as large as BOSS analysis
could. However, as predicted, due to higher resolution and signal-to-noise, smaller scales are measured. The error-bars of the flux power
used in this plot were estimated using a bootstrap covariance matrix
of the data itself (see Sec.~\ref{sec:covmat} for details) as well as the
systematic errors estimation using the method described in
Sec.~\ref{sec:results_mocks}. The XQ-100 flux power spectrum
measurements presented in this paper also agree remarkably well with the
high-redshift measurements.

\subsection{Metal flux power spectrum}
\label{sec:metal_power}

The flux power spectrum measured in this paper using the data analysis
presented in Sec.~\ref{sec:analysis} contains the power coming from
both \lya\ forest (predominantly) as well as a small contamination
from the metals.

Typically one can estimate the metal power spectrum in the QSO
spectra redwards of the \lya\ emission line, where only metal
absorption is present \citep{mcdonald05}. The absorption due to metals is coming from
mostly lower redshifts, but if unidentified it contaminates the higher
redshift \lya\ forest. It is thus further assumed that the metal
fluctuations are uncorrelated with the real \lya\ signal, and that one
can remove the effect of the metals by subtracting their power
spectrum from the measured one. If higher accuracy is desired further
corrections can be added to this approach \citep{irsic14}.

However, to measure the power spectrum redwards of the \lya\ emission
line, {\it for the same redshift range}, where flux power in the forest
  is measured, a secondary sample of lower redshift QSO spectra is
  needed. Since XQ-100 data sample only contains $100$ quasars at
  relatively high redshift, measurements of the metal power
  spectrum could only be achieved for the higher redshift bins, as
  shown in Fig.~\ref{fig:Pk_metal}. To measure the power we have
  adopted the restframe wavelength range of 
 $1268 - 1380 \,\angstrom$ in each QSO spectrum. As is
  evident from Fig.~\ref{fig:Pk_metal} the results are slightly noisy
  compared to the metal power estimated in \cite{palanque13}.

To estimate the \lya\ forest flux power for all redshifts we have
performed a simple extrapolation of the metal power spectrum
measurements. For each $k$-bin the value of metal power remains
roughly constant as a function of redshift in the measurements of
\cite{palanque13}. Using this information we averaged our $P_m(k,z)$ over the three redshift
measurements for each $k$-bin and used this as an extrapolation to
lower redshifts. This is shown in Fig.~\ref{fig:Pk_metal_zdep}. Even though
such an approximation is very rough, the value of $P_m(k,z)$ is
generally smaller or at best of the same order as the statistical
errors on our flux power spectrum measurements. 

To perform a more detailed analysis of the metal power spectrum
another sample of lower redshift quasars would be needed, or
individual metals contaminating the forest would need to be
identified. However, we believe that the results would not change
significantly and leave such a detailed analysis for future studies.

\subsection{Covariance matrix}
\label{sec:covmat}

To estimate the error-bars on the flux power spectrum the separate QSO spectra
contributions to the power spectrum were bootstrapped by assuming each
spectrum to be an independent measurement of the flux power \citep{slosar11,slosar13,irsic13}. 
We generated $1000$ bootstrapped samples of the input data-set and calculated the
corresponding bootstrap covariance matrix.

\begin{figure*}
  \centering
  \begin{tabular}{cc}
    \subfloat[${\bar F}$ variance]{
      \includegraphics[width=0.5\linewidth]{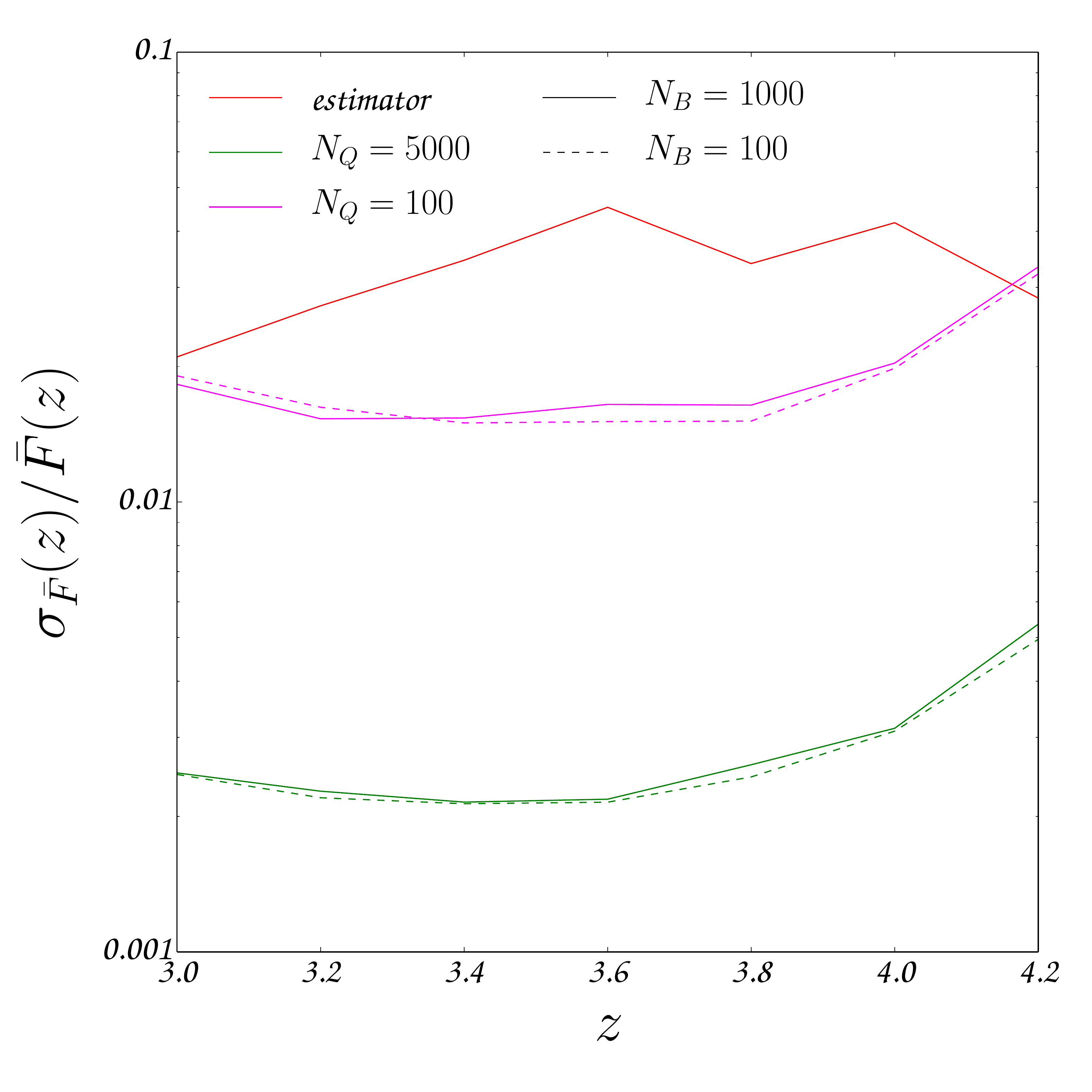}
      \label{fig:var_meanF_mocks}
    } &

    \subfloat[$P_F$ variance]{
      \includegraphics[width=0.5\linewidth]{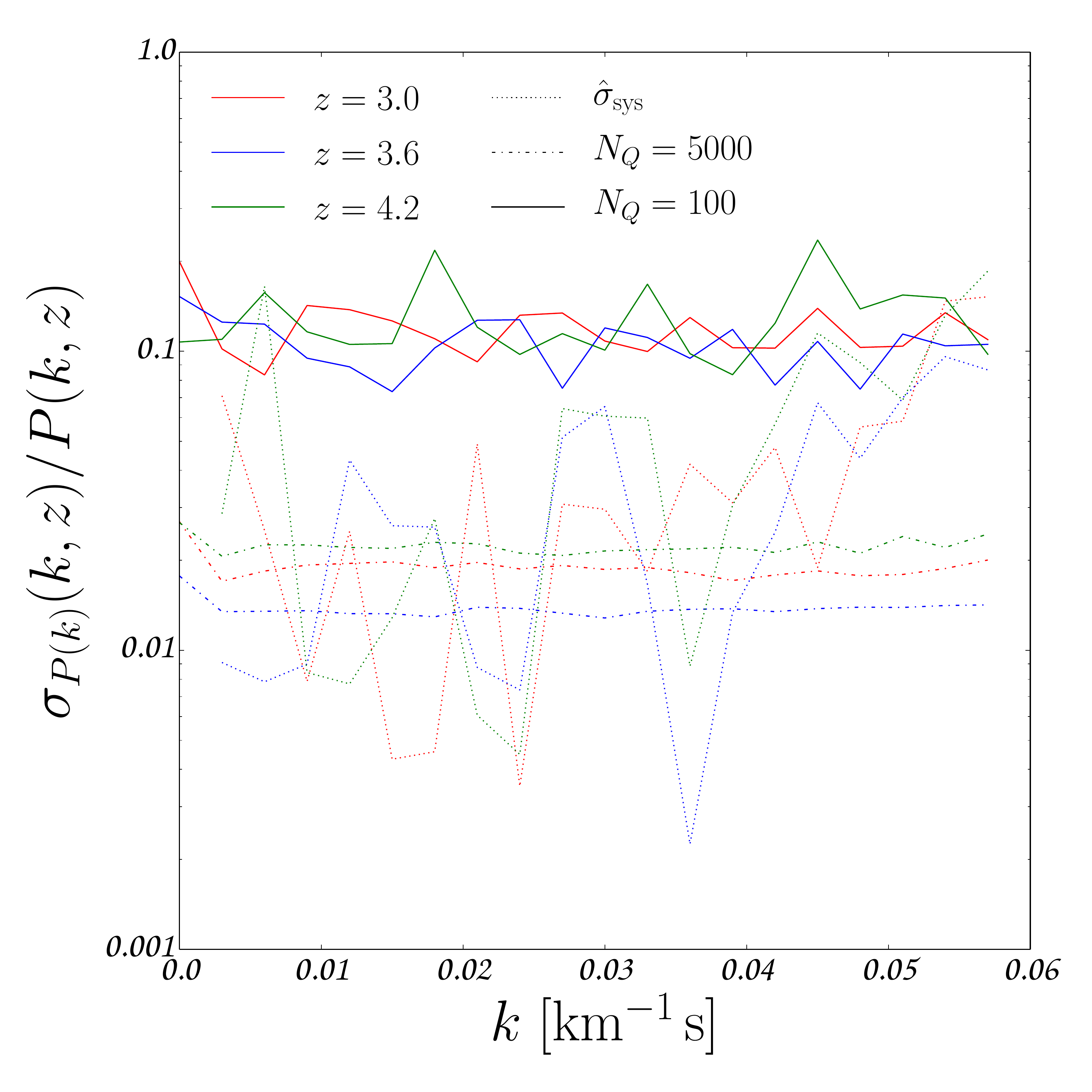}
      \label{fig:var_pk_mocks}
    } \\
  \end{tabular}
  \caption{
    {\it Left}: The estimation of relative variance on the mean flux
    measurements
    of the synthetic data sample. In red is shown the variance
    obtained through our data analysis (see Sec.~\ref{sec:meanF}),
    while magenta and green colours present the $1000$ bootstrapped
    variance of mock 100 and 5000 QSO spectra sample respectively. Dashed
    lines show the corresponding variance when only $100$ bootstrap
    samples were used. {\it Right}: The estimation of the relative variance on the flux power
    spectrum measurements of the synthetic data sample. Three colours
    correspond to three redshift bins: red - $z=3.0$, blue - $z=3.6$
    and green - $z=4.2$. full and dot-dashed lines show the results
    obtained on mock 100 and 5000 QSO spectra samples respectively
    (both with $1000$ bootstrap samples). In
    dotted lines the estimation of the systematic error is shown.
  }

\label{fig:var_mocks_ab}
\end{figure*}

The method was applied first to the synthetic data sample, for mean
flux as well as flux power spectrum
measurements. Fig.~\ref{fig:var_meanF_mocks} shows how the diagonal
elements of the bootstrapped covariance matrix (bootstrap variance)
for the mean flux changes as a function of redshift. The relative
error on the mean flux from bootstrapped samples is roughly
constant. Different line styles correspond to using $100$ or $1000$
bootstrap samples, and the differences are small. Two colour schemes
(magenta) and (green) correspond to estimating the error-bars on a
mock 100 or 5000 catalogues. The ratio between the two estimations is
exactly $\sqrt{N_Q(\text{mock 5000})}/\sqrt{N_Q(\text{mock 100})}$, meaning that
the variance scales as expected with the number of QSO spectra in the
sample ($\sim 1/\sqrt{N_Q}$). In red we plot the estimates of the mean
flux error-bars coming from the integrals over the full (signal + noise)
power spectra at each redshift bin.

\begin{figure*}
  \centering
  \begin{tabular}{cc}
    \subfloat[mock 5000]{
      \includegraphics[width=0.5\linewidth]{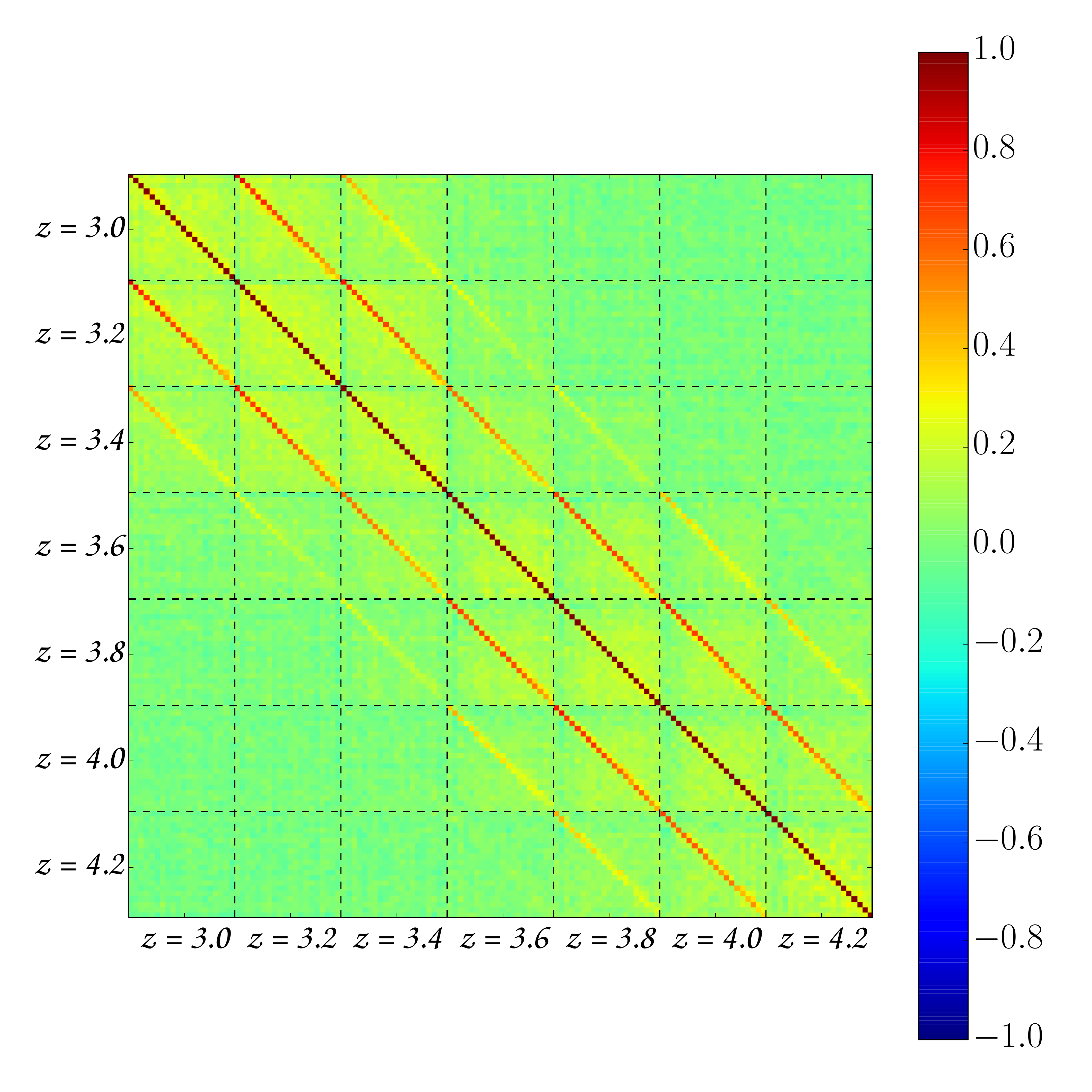}
      \label{fig:cov_pk_mocks_a}
    } &

    \subfloat[mock 100]{
      \includegraphics[width=0.5\linewidth]{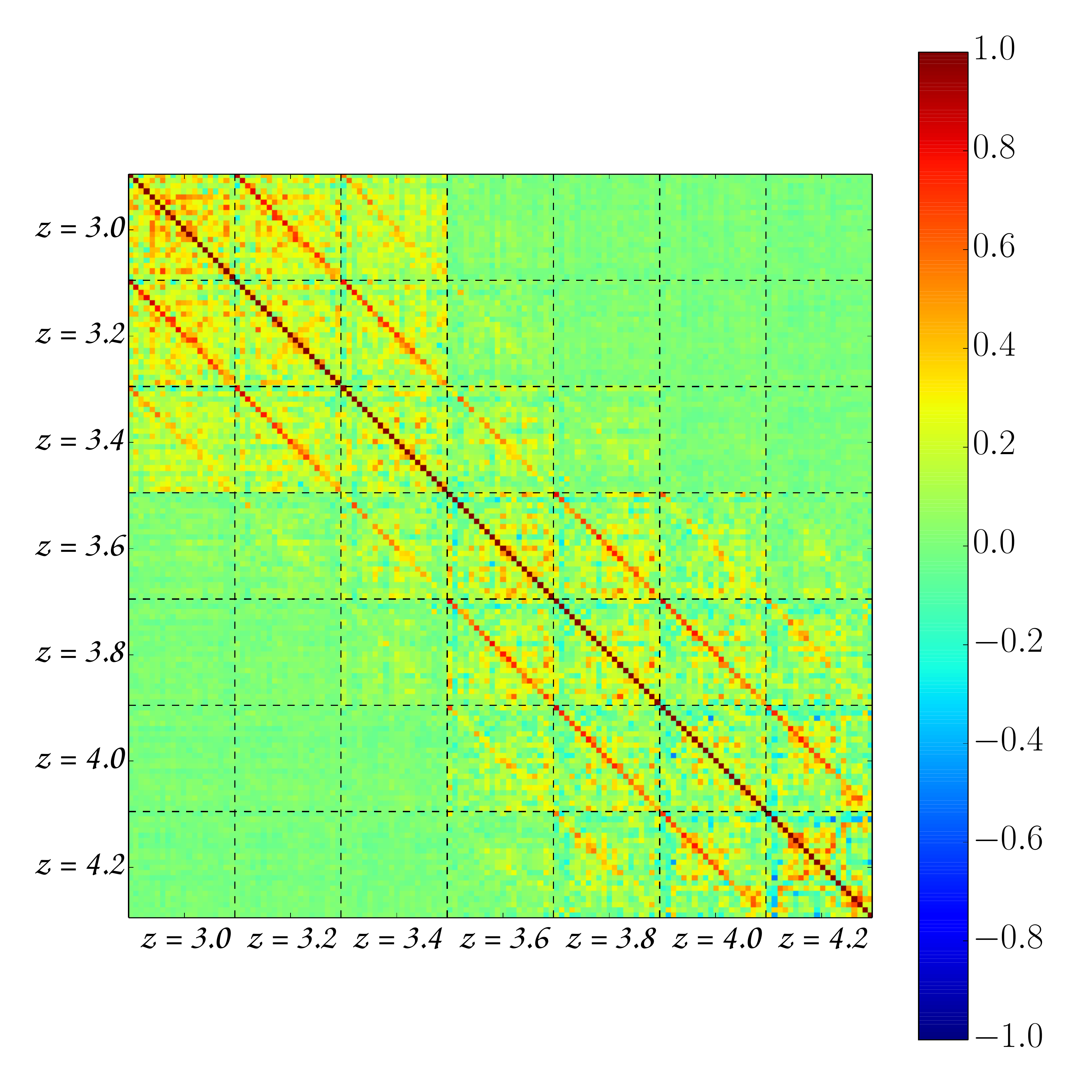}
      \label{fig:cov_pk_mocks_b}
    } \\
  \end{tabular}
  \caption{
    The error correlation matrices of the flux power spectrum
    ($C_{ij}/\sqrt{C_{ii}C_{jj}}$). Top figure
    (\ref{fig:cov_pk_mocks_a}) corresponds to the analysis done on
    $5000$ synthetic spectra, and bottom figure
    (\ref{fig:cov_pk_mocks_b}) to the analysis on only $100$
    synthetic spectra. The structure of the plot is that within each
    labeled redshift bin, the $k$-bins follow in increasing order. See
    text for details.
  }
  \label{fig:cov_pk_mocks}
\end{figure*}

Same analysis test was performed also on the flux power spectrum
variance estimation, as shown in Fig.~\ref{fig:var_pk_mocks}. Full
lines and dot-dashed lines correspond to the bootstrapped samples of
mock 100 and mock 5000 QSO spectra respectively. The scaling of the
variance holds in this case as well. In dashed lines we show the
estimation of the systematic errors on the mean flux (see
Sec.~\ref{sec:results_mocks}). 

The full bootstrap covariance matrix of the flux power spectrum is
shown in Fig.~\ref{fig:cov_pk_mocks}. The plots correspond to the
analysis done on mock 5000 (Fig.~\ref{fig:cov_pk_mocks_a}) and mock
100 (Fig.~\ref{fig:cov_pk_mocks_b}) synthetic quasar catalogues. The
covariance matrix in the plots was normalized (i.e. what is shown is
$C_{ij}/\sqrt{C_{ii}C_{jj}}$) so that the structure is readily
discernible. Within one redshift bin the correlations between
different k-bins are largely uncorrelated, with small correlation
growing from large to small scales. However the correlations between
adjacent redshift bins are quite large. This is a spurious result of
the way synthetic data are generated since up to two simulation
snapshots with successive redshift span roughly the size of one
redshift bin in the measurements.
The structure remains
basically the same (albeit noisier) when comparing the results
obtained on only a $100$ QSO spectra.

\begin{figure}
  \centering
  \includegraphics[width=1.0\linewidth]{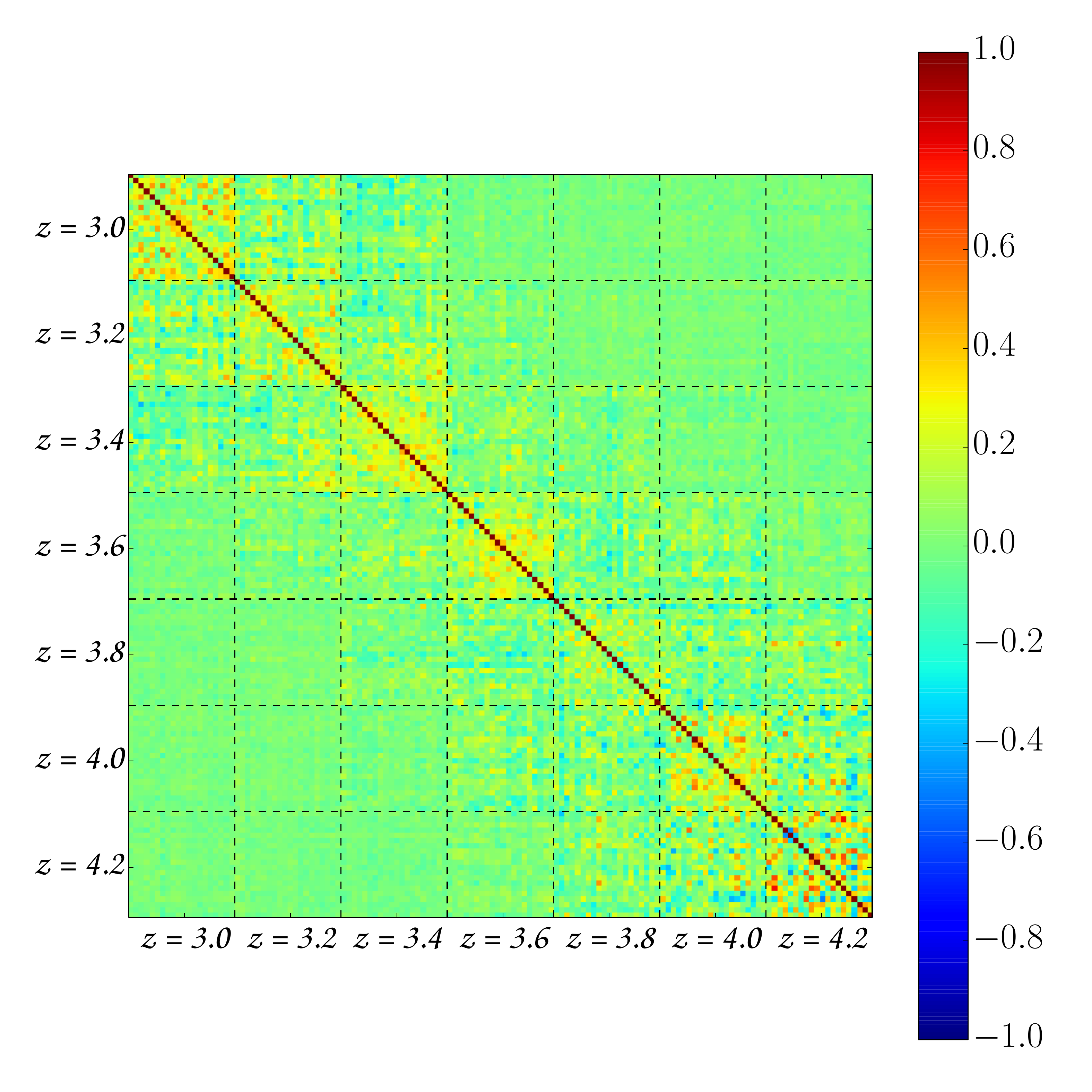}
  \caption{
    The error correlation matrix ($C_{ij}/\sqrt{C_{ii}C_{jj}}$) of the
    flux power spectrum measurements of the XQ-100 sample. See text
    for details.
  }
\label{fig:cov_pk_data}
\end{figure}

Finally, the same scheme was adopted on the XQ-100 sample, and the
results of the bootstrap covariance matrix are shown in
Fig.~\ref{fig:cov_pk_data}. The correlation matrix is somewhat noisy,
which is to be expected comparing to the analysis with varying number
of input spectra performed on the synthetic data. The correlations
with adjacent redshift bins are negligible.

Previous studies have shown, that bootstrapped covariance matrix
underestimates the variance elements of the matrix by roughly $10\%$
\citep{kim04,palanque13,busca13,slosar13,irsic13}. To compensate for
that in order to achieve a conservative estimation of the error-bars,
the full bootstrapped covariance matrix was multiplied by a factor of
$1.1$. 

\subsection{Continuum errors}
\label{sec:cont_errors}

\begin{figure}
  \centering
  \includegraphics[width=1.0\linewidth]{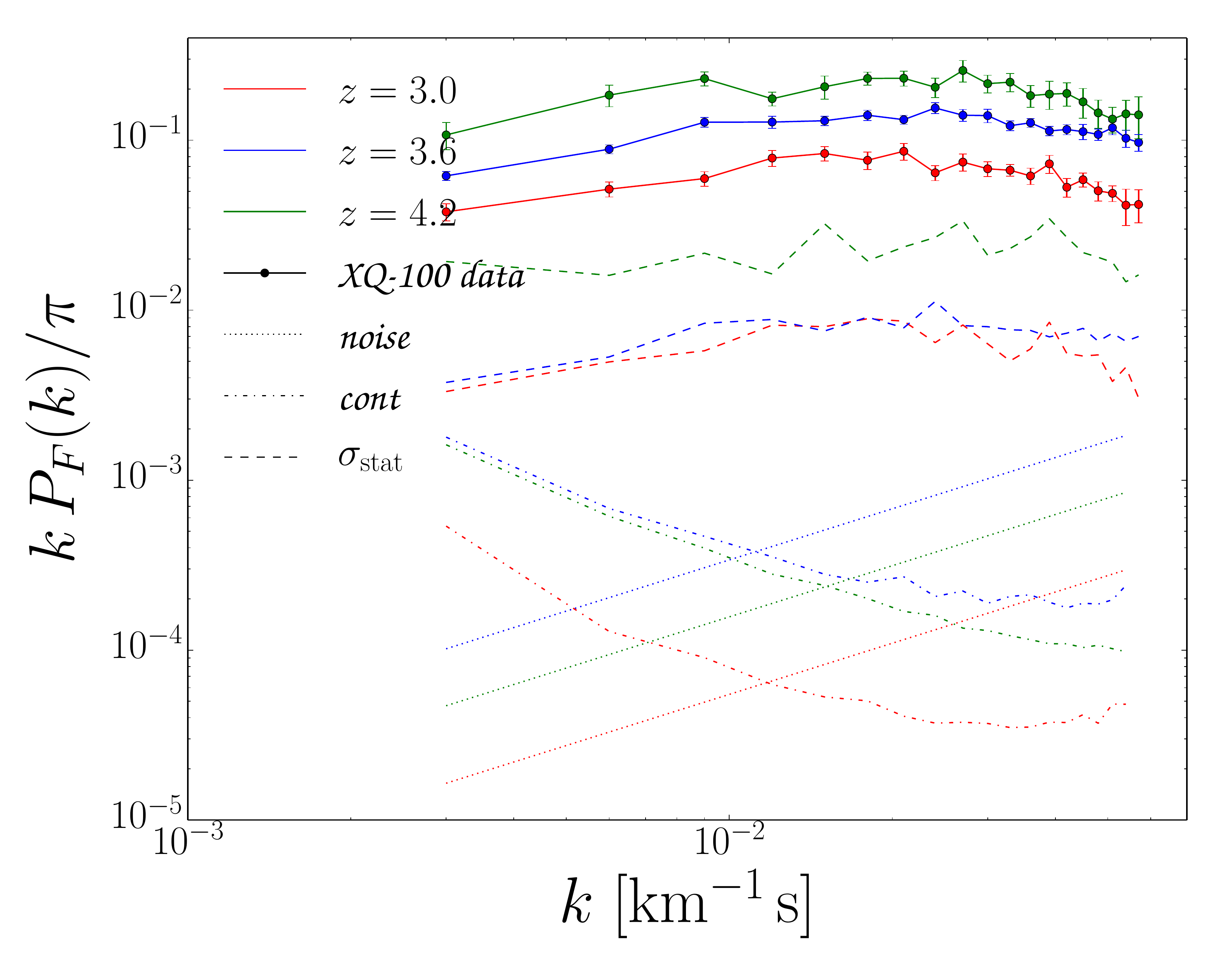}
  \caption{
    This figure shows the levels of the leaking continuum power spectrum into the
    total measured \lya\ forest power spectrum (dot-dashed
    lines). Compared to the statistical (and systematic errors)
    evaluated in the previous sections of this paper (dashed lines), uncertainties
    due to continuum fitting are small on the measurements of the
    power spectrum. The full forest flux power and the power spectrum
    of the noise are shown as a comparison (full lines and dotted
    lines, respectively). The three colours represent three
    redshift bins: $z=3.0$ (red), $z=3.6$ (blue) and $z=4.2$ (green).
  }
\label{fig:pk_cont}
\end{figure}

Since the absorption of the IGM at higher redshift becomes stronger,
it becomes hard to provide an objective estimate of the continuum
levels, due to inability to find transmission regions in the
\lya\ forest. 
{Most attempts in the
literature regarding this issue assume that either the quasar
intrinsic emission in the \lya\ forest region is unchanging from
quasar to quasar and with redshift, or they model it on a quasar-by-quasar
basis \citep{kim04,mcdonald05,palanque13,viel13,busca13,slosar13,irsic13}.

Nevertheless, the discussions and analysis on the topic in the
literature agree that a change in the normalization of the continuum
level in the \lya\ forest is perfectly degenerate with the mean
transmitted flux estimations. On the other hand, any large scale
modulations of the continuum affect the measurements of the
correlations within the forest, but when working in Fourier space,
they are confined to large scales.

To estimate the possible contamination of the continuum power leaking
into the flux power spectrum, we perform a measurements of the bare
continuum fits, as if they were representing fluctuating absorption
features of the \lya\ forest. This would be equivalent to averaging the
continua over all the lines-of-sight, to obtain an average and a
statistical description of its fluctuations. Such an approach is a
valid approximation in the limit for which we assume that all quasar
continua follow the same shape (but different normalization due to
different overall observed fluxes).
The results of this simple model are shown in
Fig.~\ref{fig:pk_cont}. The figure shows the continuum power spectra
for three different redshift bins (dot-dashed line), compared to the
levels of the statistical errors (dashed line) on the measurements of
the flux power (full line)\footnote{The systematic errors estimated in
Sec.~\ref{sec:results_mocks} are comparable to the statistical errors,
and not shown in this figure.}. The continuum power spectra show a
plateau-like feature towards smaller scales ($k > 0.01\,\skm$),
increasing in power towards large scales ($k < 0.01\,\skm$), as
expected from previous analysis.
The level of the continuum power leaking into the total forest flux
power is thus very small, indeed it is comparable to the estimated
noise power (dotted line).

While we do not use this approach in our standard analysis, it
convinces us that the systematic errors due to the continuum estimation,
that would result into increased uncertainties on very large scales
are much smaller than the statistical and systematic errors on our
measurements and can thus be neglected. We caution that this is a
simple estimation, and valid only for the data presented in this paper.

\section{Conclusions}
\label{sec:conclusion}

In this paper we have performed a \lya\ flux power spectrum analysis
on the XQ-100 sample of $100$ medium resolution, medium
signal-to-noise QSO spectra in the redshift range $3.5 < z < 4.5$ \citep{lopez16}. 
The results are shown in Fig.~\ref{fig:meanF_data} for the mean
flux measurements, in Fig.~\ref{fig:pk_data} for the flux power
spectrum measurements and in Fig.~\ref{fig:cov_pk_data} for the
estimation of the error correlations of the flux power.

The resulting mean transmitted flux is in good agreement with
previously measured mean flux by \cite{kim07} at lower redshifts. The
redshift dependence shows slight deviations from the fitting formula
in the \cite{kim07} paper at the higher redshift end, but it is still within
$1-2\sigma$ discrepancy.

Measurements of the flux power spectrum cover the range of $z=3.5-4.2$
in $7$ redshift bins and $k=0 - 0.06\,\skm$ in $20$ $k$-mode bins. The
results agree well with the expectations that despite a small sample of
QSO spectra, the higher values of spectral resolution and signal-to-noise
ratio, allow for measurements of smaller scales than a large QSO
number survey such as SDSS-III/BOSS \citep{boss13}. The total error
bars on our measurements (combined statistical and systematic) are of
the same order as those in BOSS analysis, specifically on small scales
($k > 0.01\,\skm$). At higher redshifts ($z>3.6$) our error bars are
even smaller by more than $50\%$.

In the final analysis the official (and publicly available) XQ-100
Legacy Survey continuum fits were used. To consistently measure the
mean flux (and flux power) a simultaneous measurement of the quasar
continua should be performed. However, wrong estimation of the
continuum levels would result in a slight change of normalization in
the mean flux, while any long-range modulations of the continuum are
absorbed into large scales $k$-bins in the $P_F(k)$ measurements - and
thus will not change the results on the medium to small scales this
experiment probes.

Since many QSO spectra in the sample feature
a DLA, these strong absorption system affect the flux power. 
In the current analysis we have simply removed the pixels
within $1.5$ of the DLA equivalent width around the DLA central
absorption redshift. However, with a more careful analysis DLA
component could be removed from the spectra and thus additional
wavelength ranges could be potentially added to the flux power
spectrum analysis to increase the final signal-to-noise in the
$P_F(k)$ measurements. However, since the effect on the flux power seemed to
be small and only affected large-scale $k$-bins, a simpler approach
was adopted in the final analysis of the data.

Through the use of a realistic synthetic QSO spectra sample, an
estimation of the systematic error of our data analysis was
obtained. However, for the larger part the systematic error bars are
below the statistical errors, obtained through bootstrapping the data
sample. This is valid at least in the probed $k$-mode range. At larger
scales, additional contribution to systematic errors is introduced due
to imperfect continuum fitting, while at small scales imperfect
de-convolution of the resolution/pixel width contribution introduces
significant obstacles. Last but not least, a Fourier Transform
analysis also introduces aliasing on small scales which is difficult
to correct for. For that reason such small scales (just below Nyquist
$k$-mode) were not measured in the data analysis presented in this
paper. We leave such corrections to subsequent analysis.

Due to lack of lower-redshift quasars in the XQ-100 sample, the
contaminating metal power in the \lya\ forest was only measured in
three highest redshift bins ($z=3.8-4.2$). A simple and rough
extrapolation was used to obtain an estimate of the metal power at
smaller redshifts. A separate study could be used to address this
issue. We also point out that if the metal power spectrum is measured
sufficiently accurately at all redshifts, additional second order corrections are
known to be necessary to recover the \lya\ forest flux power \citep{irsic14}.  

The results on the flux power spectrum presented in this paper have a
great potential in putting additional constraints on the cosmological
parameters, as the measurements stretch between large and small
scales, probed respectively by low-resolution large-quasar number
surveys, and a few high-resolution, high signal-to-noise QSO
spectra. The power in these intermediate scale range is sensitive to
to the small scale properties of the dark matter, as well as to
reionization epoch through the Jeans scale measurements. 

\section*{Acknowledgements}
We would like to warmly thank the ESO staff involved in the execution
of this Large Programme throughout all its phases. VI is supported by
US NSF grant AST-1514734. SL has been supported by FONDECYT grant number 1140838 and partially by PFB-06 CATA. VD, MV, SC acknowledge support from the PRIN INAF 2012 "The X-Shooter sample of 100 quasar spectra at $z \sim 3.5$: Digging into cosmology and galaxy evolution with quasar absorption lines. SLE acknowledges the receipt of an NSERC Discovery Grant. MH acknowledges support by ERC ADVANCED GRANT 320596 "The Emergence of Structure during the epoch of Reionization". The Dark Cosmology Centre is funded by the Danish National Research Foundation. MV is supported by ERC-StG "cosmoIGM". KDD is supported by an NSF AAPF fellowship awarded under NSF grant AST-1302093. JSB acknowledges the support of a Royal Society University Research Fellowship.
The hydrodynamical simulations used in this work were performed with supercomputer time awarded by the Partnership for Advanced Computing in Europe (PRACE) 8th Call. We acknowledge PRACE for awarding us access to the Curie supercomputer, based in France at the Tr\'e Grand Centre de Calcul (TGCC). This work also made use of the DiRAC High Performance Computing System (HPCS) and the COSMOS shared memory service at the University of Cambridge. These are operated on behalf of the STFC DiRAC HPC facility. This equipment is funded by BIS National E-infrastructure capital grant ST/J005673/1 and STFC grants ST/H008586/1, ST/K00333X/1. 




\bibliographystyle{mnras}
\bibliography{Bibliofile}




\appendix

\section{Table - Measured \lya\ flux power spectrum}


The last column $P_F(k,z)$ shows
  the total measured flux power spectrum, while the third column
  shows our estimate of the \lya\ forest power spectrum $P_\alpha(k,z)$, where we have
subtracted the extrapolated metal power spectrum. The second-to-last column is
measured metal power spectrum, with a dash where no data could me
measured within the XQ-100 data set. Statistical errors
($\sigma_{\text{stat}}$) were obtained using bootstrap covariance
matrix on the data. The systematic errors were obtained through
analysis on synthetic data (see Sec.~\ref{sec:results_mocks}).
The flux power spectrum and its covariance matrix can be obtained from
the following link: http://adlibitum.oats.inaf.it/XQ100survey/Data.html 

\begin{table*}
\centering
\caption{Measured \lya\ flux power spectrum from XQ-100 data
  sample. All power spectrum (and error) columns are in $[\kms]$ units. The
  scale $k$ is in $[\skm]$ units. The columns are: mean redshift and
  scale of the power spectrum bin, estimated
  \lya\ forest flux power, measured metal and total flux power, as
  well as statistical and systematic errors.}
\label{tb:pk_data}
\input{pk_table_z0}
\end{table*}

\begin{table*}
\centering
\contcaption{A table continued from the previous one}
\label{tb:pk_data:cont}
\input{pk_table_z1}
\end{table*}

\begin{table*}
\centering
\contcaption{A table continued from the previous one}
\label{tb:pk_data:cont2}
\input{pk_table_z2}
\end{table*}

\begin{table*}
\centering
\contcaption{A table continued from the previous one}
\label{tb:pk_data:cont3}
\input{pk_table_z3}
\end{table*}


\bsp	
\label{lastpage}
\end{document}

%% file: pk_table_z0.tex
\begin{tabular}{|c|c|c|c|c|c|c|}
\hline
$z$ & $k\,[\skm]$ & $P_\alpha(k,z)\,[\kms]$ & $\sigma_{\text{stat}}\,[\kms]$ & $\sigma_{\text{sys}}\,[\kms]$ & $P_m(k,z)\,[\kms]$ & $P_F(k,z)\,[\kms]$ \\ \hline\hline
3.0 & 0.003 & 39.6936 & 3.47799 & 2.96111 & - & 41.3668 \\ \hline
3.0 & 0.006 & 26.9847 & 2.59723 & 0.740328 & - & 27.7626 \\ \hline
3.0 & 0.009 & 20.7667 & 2.01472 & 0.197852 & - & 21.6864 \\ \hline
3.0 & 0.012 & 20.5633 & 2.13739 & 0.518419 & - & 21.3302 \\ \hline
3.0 & 0.015 & 17.4999 & 1.67476 & 0.077641 & - & 18.1371 \\ \hline
3.0 & 0.018 & 13.3093 & 1.55476 & 0.0705116 & - & 13.6733 \\ \hline
3.0 & 0.021 & 12.8818 & 1.28926 & 0.677582 & - & 13.1812 \\ \hline
3.0 & 0.024 & 8.42079 & 0.843842 & 0.0403608 & - & 8.85882 \\ \hline
3.0 & 0.027 & 8.65179 & 0.952071 & 0.297344 & - & 9.03964 \\ \hline
3.0 & 0.03 & 7.11185 & 0.663833 & 0.251926 & - & 7.34621 \\ \hline
3.0 & 0.033 & 6.34019 & 0.478553 & 0.14305 & - & 6.52495 \\ \hline
3.0 & 0.036 & 5.38066 & 0.516292 & 0.293888 & - & 5.68213 \\ \hline
3.0 & 0.039 & 5.84972 & 0.684576 & 0.190915 & - & 6.11011 \\ \hline
3.0 & 0.042 & 3.9562 & 0.41663 & 0.263036 & - & 4.11763 \\ \hline
3.0 & 0.045 & 4.0884 & 0.375191 & 0.0865709 & - & 4.21108 \\ \hline
3.0 & 0.048 & 3.299 & 0.357229 & 0.232406 & - & 3.47645 \\ \hline
3.0 & 0.051 & 3.00056 & 0.234432 & 0.217284 & - & 3.15771 \\ \hline
3.0 & 0.054 & 2.41408 & 0.268641 & 0.517127 & - & 2.55463 \\ \hline
3.0 & 0.057 & 2.30919 & 0.167658 & 0.483684 & - & 2.38873 \\ \hline
3.2 & 0.003 & 50.5538 & 3.76044 & 3.48715 & - & 52.227 \\ \hline
3.2 & 0.006 & 36.6106 & 2.45683 & 4.6167 & - & 37.3885 \\ \hline
3.2 & 0.009 & 29.5313 & 1.86716 & 0.141733 & - & 30.451 \\ \hline
3.2 & 0.012 & 23.1562 & 1.58399 & 0.42266 & - & 23.9231 \\ \hline
3.2 & 0.015 & 17.56 & 1.14134 & 0.822192 & - & 18.1972 \\ \hline
3.2 & 0.018 & 15.714 & 0.856297 & 1.42563 & - & 16.078 \\ \hline
3.2 & 0.021 & 14.8417 & 1.04861 & 0.526579 & - & 15.1411 \\ \hline
3.2 & 0.024 & 12.6287 & 1.02804 & 0.626563 & - & 13.0667 \\ \hline
3.2 & 0.027 & 10.3106 & 0.762546 & 1.31213 & - & 10.6985 \\ \hline
3.2 & 0.03 & 9.74955 & 0.807138 & 0.638237 & - & 9.98391 \\ \hline
3.2 & 0.033 & 8.73696 & 0.571104 & 1.07065 & - & 8.92172 \\ \hline
3.2 & 0.036 & 7.87109 & 0.607356 & 0.299586 & - & 8.17256 \\ \hline
3.2 & 0.039 & 7.55867 & 0.554116 & 0.14941 & - & 7.81906 \\ \hline
3.2 & 0.042 & 6.41257 & 0.379081 & 0.193592 & - & 6.574 \\ \hline
3.2 & 0.045 & 5.253 & 0.384992 & 0.0941316 & - & 5.37568 \\ \hline
3.2 & 0.048 & 4.63914 & 0.336089 & 0.127258 & - & 4.81659 \\ \hline
3.2 & 0.051 & 4.2735 & 0.306388 & 0.295822 & - & 4.43065 \\ \hline
3.2 & 0.054 & 3.775 & 0.236653 & 0.220056 & - & 3.91555 \\ \hline
3.2 & 0.057 & 3.25205 & 0.183355 & 0.281402 & - & 3.33159 \\ \hline
\end{tabular}

%% file: pk_table_z1.tex
\begin{tabular}{|c|c|c|c|c|c|c|}
\hline
$z$ & $k\,[\skm]$ & $P_\alpha(k,z)\,[\kms]$ & $\sigma_{\text{stat}}\,[\kms]$ & $\sigma_{\text{sys}}\,[\kms]$ & $P_m(k,z)\,[\kms]$ & $P_F(k,z)\,[\kms]$ \\ \hline\hline
3.4 & 0.003 & 54.6488 & 3.67166 & 1.31521 & - & 56.322 \\ \hline
3.4 & 0.006 & 45.1101 & 2.37959 & 2.05999 & - & 45.888 \\ \hline
3.4 & 0.009 & 33.6866 & 1.90515 & 1.59619 & - & 34.6063 \\ \hline
3.4 & 0.012 & 29.4042 & 1.83353 & 1.34494 & - & 30.1711 \\ \hline
3.4 & 0.015 & 22.2285 & 1.22494 & 0.165127 & - & 22.8657 \\ \hline
3.4 & 0.018 & 21.4314 & 1.21646 & 1.31261 & - & 21.7954 \\ \hline
3.4 & 0.021 & 18.3216 & 1.12336 & 0.561181 & - & 18.621 \\ \hline
3.4 & 0.024 & 16.861 & 1.0669 & 0.458334 & - & 17.299 \\ \hline
3.4 & 0.027 & 13.1393 & 0.752371 & 0.199968 & - & 13.5272 \\ \hline
3.4 & 0.03 & 12.1581 & 0.747101 & 1.47473 & - & 12.3925 \\ \hline
3.4 & 0.033 & 10.8306 & 0.739907 & 0.217761 & - & 11.0154 \\ \hline
3.4 & 0.036 & 9.94063 & 0.65352 & 0.439429 & - & 10.2421 \\ \hline
3.4 & 0.039 & 8.85191 & 0.541132 & 0.149585 & - & 9.1123 \\ \hline
3.4 & 0.042 & 7.06202 & 0.441117 & 0.406511 & - & 7.22345 \\ \hline
3.4 & 0.045 & 7.21777 & 0.520931 & 0.0199403 & - & 7.34045 \\ \hline
3.4 & 0.048 & 6.5484 & 0.394166 & 0.0315864 & - & 6.72585 \\ \hline
3.4 & 0.051 & 5.54113 & 0.357655 & 0.00463702 & - & 5.69828 \\ \hline
3.4 & 0.054 & 5.33921 & 0.318303 & 0.279183 & - & 5.47976 \\ \hline
3.4 & 0.057 & 4.79408 & 0.294146 & 0.180972 & - & 4.87362 \\ \hline
3.6 & 0.003 & 64.6285 & 3.93553 & 0.572105 & - & 66.3017 \\ \hline
3.6 & 0.006 & 46.3763 & 2.77871 & 0.391996 & - & 47.1542 \\ \hline
3.6 & 0.009 & 44.561 & 2.92414 & 0.369775 & - & 45.4807 \\ \hline
3.6 & 0.012 & 33.4982 & 2.30848 & 1.55865 & - & 34.2651 \\ \hline
3.6 & 0.015 & 27.2763 & 1.57973 & 0.76344 & - & 27.9135 \\ \hline
3.6 & 0.018 & 24.5006 & 1.58857 & 0.619505 & - & 24.8646 \\ \hline
3.6 & 0.021 & 19.7668 & 1.17869 & 0.187915 & - & 20.0662 \\ \hline
3.6 & 0.024 & 20.2644 & 1.47368 & 0.144347 & - & 20.7024 \\ \hline
3.6 & 0.027 & 16.3306 & 0.943949 & 0.911005 & - & 16.7185 \\ \hline
3.6 & 0.03 & 14.6182 & 0.837248 & 1.03125 & - & 14.8526 \\ \hline
3.6 & 0.033 & 11.5936 & 0.729936 & 0.218077 & - & 11.7784 \\ \hline
3.6 & 0.036 & 11.0542 & 0.663244 & 0.0259277 & - & 11.3557 \\ \hline
3.6 & 0.039 & 9.13545 & 0.561005 & 0.137402 & - & 9.39584 \\ \hline
3.6 & 0.042 & 8.65139 & 0.547743 & 0.221311 & - & 8.81282 \\ \hline
3.6 & 0.045 & 7.84233 & 0.545822 & 0.572906 & - & 7.96501 \\ \hline
3.6 & 0.048 & 7.07895 & 0.429794 & 0.330696 & - & 7.2564 \\ \hline
3.6 & 0.051 & 7.29084 & 0.450913 & 0.471224 & - & 7.44799 \\ \hline
3.6 & 0.054 & 5.97605 & 0.38174 & 0.574253 & - & 6.1166 \\ \hline
3.6 & 0.057 & 5.35158 & 0.386213 & 0.466262 & - & 5.43112 \\ \hline
\end{tabular}

%% file: pk_table_z2.tex
\begin{tabular}{|c|c|c|c|c|c|c|}
\hline
$z$ & $k\,[\skm]$ & $P_\alpha(k,z)\,[\kms]$ & $\sigma_{\text{stat}}\,[\kms]$ & $\sigma_{\text{sys}}\,[\kms]$ & $P_m(k,z)\,[\kms]$ & $P_F(k,z)\,[\kms]$ \\ \hline\hline
3.8 & 0.003 & 94.9659 & 4.78048 & 3.15856 & 2.20798 & 97.1739 \\ \hline
3.8 & 0.006 & 64.7637 & 5.17604 & 1.52262 & 0.618829 & 65.3825 \\ \hline
3.8 & 0.009 & 51.1572 & 3.0288 & 0.770617 & 1.01263 & 52.1698 \\ \hline
3.8 & 0.012 & 41.4319 & 3.02359 & 1.43161 & 0.710195 & 42.1421 \\ \hline
3.8 & 0.015 & 35.5927 & 2.77421 & 0.320939 & 0.757387 & 36.3501 \\ \hline
3.8 & 0.018 & 31.0847 & 2.22233 & 0.209592 & 0.44708 & 31.5318 \\ \hline
3.8 & 0.021 & 26.1317 & 1.57351 & 0.170671 & 0.3575 & 26.4892 \\ \hline
3.8 & 0.024 & 20.8432 & 1.57842 & 0.266877 & 0.55192 & 21.3951 \\ \hline
3.8 & 0.027 & 18.3734 & 1.34587 & 0.00462213 & 0.558363 & 18.9318 \\ \hline
3.8 & 0.03 & 15.6996 & 1.15623 & 0.621457 & 0.335736 & 16.0353 \\ \hline
3.8 & 0.033 & 14.5548 & 0.924196 & 0.209155 & 0.192244 & 14.747 \\ \hline
3.8 & 0.036 & 11.8421 & 0.884394 & 0.360085 & 0.433281 & 12.2754 \\ \hline
3.8 & 0.039 & 10.5212 & 0.661361 & 0.370649 & 0.301957 & 10.8232 \\ \hline
3.8 & 0.042 & 10.3815 & 0.811933 & 0.326826 & 0.18257 & 10.5641 \\ \hline
3.8 & 0.045 & 9.4877 & 0.582263 & 0.466667 & 0.15454 & 9.64224 \\ \hline
3.8 & 0.048 & 8.20416 & 0.531247 & 0.412172 & 0.211622 & 8.41578 \\ \hline
3.8 & 0.051 & 6.52915 & 0.466894 & 0.355527 & 0.193798 & 6.72295 \\ \hline
3.8 & 0.054 & 6.01328 & 0.467307 & 0.530194 & 0.195322 & 6.2086 \\ \hline
3.8 & 0.057 & 5.25984 & 0.430496 & 0.592233 & 0.0734851 & 5.33333 \\ \hline
4.0 & 0.003 & 111.399 & 10.3519 & 3.8603 & 0.8259 & 112.225 \\ \hline
4.0 & 0.006 & 68.0376 & 5.04346 & 2.3163 & 0.517393 & 68.555 \\ \hline
4.0 & 0.009 & 52.8667 & 5.3455 & 1.82327 & 0.570465 & 53.4372 \\ \hline
4.0 & 0.012 & 48.6669 & 4.45016 & 4.65906 & 0.535946 & 49.2028 \\ \hline
4.0 & 0.015 & 41.833 & 3.42083 & 1.51375 & 0.394923 & 42.2279 \\ \hline
4.0 & 0.018 & 29.9528 & 2.45372 & 0.581078 & 0.222861 & 30.1757 \\ \hline
4.0 & 0.021 & 33.2897 & 2.75049 & 0.120818 & 0.18742 & 33.4771 \\ \hline
4.0 & 0.024 & 28.6796 & 2.3593 & 0.771985 & 0.3025 & 28.9821 \\ \hline
4.0 & 0.027 & 24.4995 & 2.33264 & 0.69748 & 0.242358 & 24.7419 \\ \hline
4.0 & 0.03 & 21.8006 & 1.87943 & 0.592556 & 0.124732 & 21.9253 \\ \hline
4.0 & 0.033 & 16.7224 & 1.48355 & 1.90312 & 0.118568 & 16.841 \\ \hline
4.0 & 0.036 & 14.9103 & 1.3448 & 0.560797 & 0.159498 & 15.0698 \\ \hline
4.0 & 0.039 & 14.4875 & 1.30919 & 0.113916 & 0.175863 & 14.6634 \\ \hline
4.0 & 0.042 & 12.0117 & 1.05231 & 1.13926 & 0.0995424 & 12.1112 \\ \hline
4.0 & 0.045 & 10.2184 & 0.786477 & 1.02756 & 0.0734334 & 10.2918 \\ \hline
4.0 & 0.048 & 9.49697 & 0.662276 & 0.727909 & 0.126871 & 9.62384 \\ \hline
4.0 & 0.051 & 8.66586 & 0.707745 & 0.91366 & 0.113339 & 8.7792 \\ \hline
4.0 & 0.054 & 6.85466 & 0.612431 & 1.1111 & 0.0813378 & 6.936 \\ \hline
4.0 & 0.057 & 7.21086 & 0.630264 & 1.21433 & 0.0649048 & 7.27576 \\ \hline
\end{tabular}

%% file: pk_table_z3.tex
\begin{tabular}{|c|c|c|c|c|c|c|}
\hline
$z$ & $k\,[\skm]$ & $P_\alpha(k,z)\,[\kms]$ & $\sigma_{\text{stat}}\,[\kms]$ & $\sigma_{\text{sys}}\,[\kms]$ & $P_m(k,z)\,[\kms]$ & $P_F(k,z)\,[\kms]$ \\ \hline\hline
4.2 & 0.003 & 112.535 & 20.2444 & 3.0241 & 1.98573 & 114.521 \\ \hline
4.2 & 0.006 & 96.4383 & 8.39488 & 11.2881 & 1.19746 & 97.6358 \\ \hline
4.2 & 0.009 & 80.4612 & 7.55641 & 0.560461 & 1.17602 & 81.6372 \\ \hline
4.2 & 0.012 & 45.9185 & 4.27574 & 0.427649 & 1.05457 & 46.9731 \\ \hline
4.2 & 0.015 & 43.2432 & 6.7416 & 0.597927 & 0.759341 & 44.0025 \\ \hline
4.2 & 0.018 & 40.3104 & 3.4052 & 1.07437 & 0.421928 & 40.7323 \\ \hline
4.2 & 0.021 & 34.65 & 3.52365 & 0.21666 & 0.353345 & 35.0033 \\ \hline
4.2 & 0.024 & 26.8387 & 3.50562 & 0.139405 & 0.459671 & 27.2984 \\ \hline
4.2 & 0.027 & 29.9342 & 3.90649 & 1.88457 & 0.362843 & 30.297 \\ \hline
4.2 & 0.03 & 22.545 & 2.20596 & 1.55529 & 0.242615 & 22.7876 \\ \hline
4.2 & 0.033 & 20.9133 & 2.19605 & 1.34463 & 0.243462 & 21.1568 \\ \hline
4.2 & 0.036 & 15.9715 & 2.35841 & 0.167869 & 0.311622 & 16.2831 \\ \hline
4.2 & 0.039 & 15.0499 & 2.78053 & 0.550652 & 0.303354 & 15.3533 \\ \hline
4.2 & 0.042 & 14.0982 & 2.00849 & 0.926213 & 0.202175 & 14.3004 \\ \hline
4.2 & 0.045 & 11.7465 & 1.52352 & 1.77379 & 0.140056 & 11.8866 \\ \hline
4.2 & 0.048 & 9.48362 & 1.33956 & 1.18794 & 0.193858 & 9.67748 \\ \hline
4.2 & 0.051 & 8.204 & 1.17976 & 0.772654 & 0.164307 & 8.36831 \\ \hline
4.2 & 0.054 & 8.31223 & 0.855154 & 1.44071 & 0.144993 & 8.45722 \\ \hline
4.2 & 0.057 & 7.77377 & 0.888949 & 1.9529 & 0.100243 & 7.87401 \\ \hline
\end{tabular}